\documentclass[singlecolumn,showpacs,preprintnumbers,amsmath,amssymb,floatfix]{revtex4}

\usepackage{graphicx}
\usepackage{dcolumn}
\usepackage{bm}

\newcommand{\beq}{\begin{equation}}
\newcommand{\eeq}{\end{equation}}
\newcommand{\bey}{\begin{eqnarray}}
\newcommand{\eey}{\end{eqnarray}}

\begin{document}

\title{Charged anisotropic matter with linear or nonlinear equation of state}

\author{Victor Varela\footnote{victor.varela.abdn@gmail.com}}
\affiliation{Institute of Mathematics, King's College, University
of Aberdeen, Aberdeen AB24 3UE, UK}

\author{Farook Rahaman\footnote{farook\_rahaman@yahoo.com}}
\affiliation{Department of Mathematics, Jadavpur University, Kolkata 700 032, West Bengal, India}

\author{Saibal Ray\footnote{saibal@iucaa.ernet.in}}
\affiliation{Department of Physics, Government College of Engineering
\& Ceramic Technology, Kolkata 700 010, West Bengal, India}

\author{Koushik Chakraborty\footnote{E-mail: kchakraborty28@yahoo.com}}
\affiliation{Department of Physics, Government Training College,
Hooghly 712103, India}

\author{Mehedi Kalam\footnote{E-mail: mehedikalam@yahoo.co.in}}
\affiliation{Department of Physics, Netaji Nagar College for
Women, Kolkata 700092, India}

\date{\today}

\begin{abstract}
Ivanov pointed out substantial analytical difficulties associated
with self-gravitating, static, isotropic fluid spheres when
pressure explicitly depends on matter density. Simplifications
achieved with the introduction of electric charge were noticed as
well. We deal with self-gravitating,  charged, anisotropic fluids
and get even more flexibility in solving the Einstein-Maxwell
equations.  In order to discuss analytical solutions we extend
Krori and Barua's method to include pressure  anisotropy and linear
or non-linear equations of state. The field equations are reduced
to a system of three algebraic equations for the anisotropic
pressures as well as matter and electrostatic energy densities.
Attention is paid to compact sources characterized by positive
matter density and positive radial pressure. Arising solutions
satisfy the energy conditions of general relativity. Spheres
with vanishing net charge contain fluid elements with unbounded proper
charge density located at the fluid-vacuum interface. Notably the
electric force acting on these fluid elements is finite, although
the acting electric field is zero. Net charges can be huge
($10^{19}\,C$) and maximum electric field intensities are very
large ($10^{23}-10^{24}\,statvolt/cm$) even in the case of zero
net charge. Inward-directed fluid forces caused by pressure
anisotropy may allow equilibrium configurations with larger net
charges and electric field intensities than those found in studies
of charged isotropic fluids. Links of these results with charged
strange quark stars as well as models of dark matter including
massive charged particles are highlighted. The van der Waals
equation of state leading to matter densities constrained by cubic
polynomial equations is briefly considered. The fundamental
question of stability is left open.
\end{abstract}


\maketitle

\section{Introduction}

Self-gravitating fluid models are essential to many applications
of general relativity, ranging from the historically important
classical models of elementary particles to current computational
and observational problems in astrophysics and cosmology. These
systems are usually characterized by sets of physical variables
which outnumber the independent field equations. Consistent
problems can be posed if additional restrictions on the variables
are specified, which may take the form of equations of state
(EOS).

Using an EOS to describe a self-gravitating fluid has important
consequences when it comes to solving the field equations. For
example, Ivanov \cite{iva1} has pointed out that finding
analytical solutions in the static, spherically symmetric,
uncharged case of a perfect fluid with linear EOS is an extremely
difficult problem.

Interestingly, analytical difficulties may alleviate with
increasing physical complexity. This situation
has been illustrated by Sharma and Maharaj \cite{sm1} in the
case of a static, spherically symmetric,
uncharged anisotropic fluid. These authors chose a particular
mass function to reduce and easily solve
the system of field equations combined with a linear EOS.

The surprising simplification of the field equations for a charged
perfect fluid satisfying a linear EOS was discussed by Ivanov
\cite{iva1}, who showed how to reduce the system involving the
most general linear EOS to a linear differential equation for one
metric component. However Ivanov also pointed out that the use of
a polytropic EOS leads to non-integrable equations.

Electrically charged fluids with anisotropic pressures constitute
the next level of physical complexity.

Charged, self-gravitating anisotropic fluid spheres have been
investigated in general relativity since the pioneering work of
Bonnor \cite{bon}. Recently, this type of charged matter has been
considered by Horvat, Iliji{\'c} and Marunovi{\'c} in studies of
gravastars \cite{gs1}. Models with prescribed EOS remain
relatively unexplored.

Motivated by MIT bag models of strange stars, Thirukkanesh and
Maharaj \cite{tm1} derived solutions for charged anisotropic
fluids with linear EOS from specific choices of one metric
function and the electrostatic energy density. Their method - an
extension of the procedure presented in \cite{sm1} for uncharged
anisotropic fluids - provides motivation for the completely
different approach developed in this paper.

Current dark matter and dark energy models are associated with
nonlinear EOS. Density perturbations may lead to fluid nucleation
characterized by heterogeneous matter densities and pressures.
Assuming specific EOS, bounded static fluid distributions and
asymptotically flat spacetimes, we aim to study the arising
self-gravitating objects. We insist on analytical solutions and
note that prescribed EOS may cause serious trouble. However, we
find that the convenient choice of a ``frozen" internal metric
reduces the Einstein-charged fluid equations to a system of linear
algebraic equations for matter and electrostatic energy densities
as well as anisotropic pressures. In a way, we are led to the
simplest solution method available for this type of source with
prescribed EOS and asymptotically flat spacetime. Models based on
linear and nonlinear EOS are completely solved following
essentially the same procedure. Remarkably, our approach uncovers
the effects of different EOS on hydrostatic and electrical
variables without interfering changes of internal metric (apart
from adjustable numerical factors depending on junction
conditions). This treatment offers a fresh view of the
relationships among EOS, charge distributions and pressure
anisotropy.

In Section II we write the field equations and briefly review the
work of Thirukkanesh and Maharaj. Also, the original Krori and
Barua solution method is generalized to deal with charged
anisotropic sources with regular interiors. Our approach to linear
and nonlinear EOS leading to models with positive definite matter
density is presented in Section III. In Section IV we discuss
junction conditions and reformulate our procedure in terms of
dimensionless quantities. Section V includes detailed analysis of
models with positive matter density and positive radial pressure,
paying special attention to conditions for physical acceptability
as well as equilibrium conditions and the values of key physical
parameters in Gaussian-cgs units. In Section VI we speculate on
the origin of charge in models with linear and nonlinear EOS, and
check our solution method against a more complicated EOS
describing quintessence stars. Finally we suggest avenues for
further research involving stability analysis and gravitational
collapse.

\section{CHARGED ANISOTROPIC MATTER}

The starting point is the static, spherically symmetric line
element represented in curvature coordinates. It reads
\begin{equation}
    ds^2=e^{\nu} dt^2-e^{\lambda} dr^2-r^2 d\theta^2-r^2\sin^2\theta d\phi^2, \label{le}
\end{equation}
where $\nu=\nu(r)$ and $\lambda=\lambda(r)$. For the static,
charged source with density $\rho=\rho(r)$, radial pressure
$p_r=p_r(r)$, tangential pressure $p_t=p_t(r)$, proper charge
density $\sigma=\sigma(r)$ and electric field $E=E(r)$ the
Einstein-Maxwell
 (EM) equations take the form
\begin{equation}
    8\pi\rho+E^2=e^{-\lambda}\left(\frac{\lambda^\prime}{r}-\frac{1}{r^2}\right)+\frac{1}{r^2}, \label{em1}
\end{equation}
\begin{equation}
    8\pi p_r-E^2=e^{-\lambda}\left(\frac{\nu^\prime}{r}+\frac{1}{r^2}\right)-\frac{1}{r^2}, \label{em2}
\end{equation}
\begin{equation}
    8\pi p_t+E^2=\frac{e^{-\lambda}}{2}\left( \nu^{\prime\prime}+\frac{\nu^{\prime 2}}{2}+\frac{\nu^{\prime}
    -{\lambda^{\prime}}}{r}-\frac{\nu^{\prime}\lambda^{\prime}}{2}\right), \label{em3}
\end{equation}
\begin{equation}
  \sigma=\frac{e^{-\frac{\lambda}{2}}}{4\pi r^2}\left( r^2 E \right)^\prime, \label{em4}
\end{equation}
where the primes denote differentiation with respect to $r$, and
geometrized units $(G=c=1)$ are employed.

Equations (\ref{em1})-(\ref{em4}) are invariant under the
transformation $E\,\rightarrow\,-E$, $\sigma\,
\rightarrow\,-\sigma$. In this work we exclusively deal with the
positive square root of $E^2$.

With our choice of radial coordinate the metric function
$e^{\lambda}$ and electrostatic energy density $E^2$ assumed in
\cite{tm1} take the forms
\begin{equation}
e^{\lambda}=\frac{1+a r^2}{1+\left(a-b\right)r^2}, \label{tm1}
\end{equation}
\begin{equation}
E^2=\frac{k\left(3+ar^2\right)}{\left(1+ar^2 \right)^2},
\label{tm2}
\end{equation}
where $a$, $b$, $k$ are arbitrary constants.

The combination of (\ref{tm1}) and (\ref{tm2}) with (\ref{em1}) yields
\begin{equation}
\rho=\frac{(b-k)\left(3+ar^2\right)}{8\pi\left(1+ar^2 \right)^2}. \label{tmrho}
\end{equation}
Hence $\rho$ and $E^2$ are proportional when $k \neq 0$ and $k
\neq b$. This expression for $\rho(r)$ is joined to a linear EOS
to provide $p_r(r)$. The explicit form of $p_r(r)$ together with
(\ref{tm1}), (\ref{tm2}) and (\ref{em2}) imply a linear
differential equation for $\nu(r)$ which is analytically solved.
Finally, when (\ref{tm1}), (\ref{tm2}) and the explicit form of
$\nu(r)$ are substituted into (\ref{em3}) we get the corresponding
expression for $p_t(r)$.

The above procedure leads to analytical solutions which depend on
a number of free parameters. The analysis presented in \cite{tm1}
considers sensible choices for these parameters in the context of
charged stars. However we find a singularity in the charge
distribution at $r=0$ when (\ref{tm1}) and (\ref{tm2}) are
combined with (\ref{em4}).

From (\ref{tm1}) we see that $\lambda(r)$ satisfies
$\lambda(0)=0$. Assuming $k>0$, (\ref{em4}) implies
\begin{equation}
\sigma(r)\approx\frac{\sqrt{3k}}{2\pi}\frac{1}{r}   \label{tmsigma}
\end{equation}
for small $r$, which diverges at $r=0$. The electric field
associated with (\ref{tm2}) does not vanish at $r=0$. This choice
for $E(r)$ prevents the regularity of the charge distribution at
the centre of the sphere.

In our view, the vanishing of the electric field at the center of
a spherically symmetric charge distribution should be a condition
for physical relevance of the solution. Furthermore the
significance of charge density singularities is unclear and we aim
to investigate their possible occurrence in models satisfying
$E(0)=0$. To this end, we extend Krori and Barua's approach to
charged isotropic fluid sources \cite{kb} and deal with charged
anisotropic sources with prescribed EOS.

Krori and Barua (KB) constructed singularity-free models of
static, charged perfect fluids with metric (\ref{le}) given by
\begin{equation}
\lambda=Ar^2,   \label{kb1}
\end{equation}
\begin{equation}
\nu=Br^2+C,  \label{kb2}
\end{equation}
where $A$, $B$ and $C$ are constants. This internal metric
satisfies the conditions for regularity at $r=0$ discussed by Lake
and Musgrave \cite{lamu}. As a consequence of this choice,
equations (\ref{em1})-(\ref{em3}) with $p_r=p_t=p$ were reduced to
a system of three linear algebraic equations for  $\rho$, $p$ and
$E^2$. Furthermore $\sigma$ was obtained combining (\ref{em4})
with the chosen square root of $E^2$ and the assumed form of
$\lambda$.

We re-write the field equations as
\begin{equation}
8\pi\rho+E^2=f(r), \label{fe1}
\end{equation}
\begin{equation}
8\pi p_r-E^2=h(r),  \label{fe2}
\end{equation}
\begin{equation}
8\pi p_t+E^2=j(r),  \label{fe3}
\end{equation}
where $f(r), h(r), j(r)$ are determined by the right sides of
(\ref{em1})-(\ref{em3}), (\ref{kb1}) and (\ref{kb2}) namely
\begin{equation}
    f(r)=e^{-Ar^2}\left(2A-\frac{1}{r^2}\right)+\frac{1}{r^2}, \label{ff}
\end{equation}
\begin{equation}
    h(r)=e^{-Ar^2}\left(2B+\frac{1}{r^2}\right)-\frac{1}{r^2}, \label{fh}
\end{equation}
\begin{equation}
    j(r)=e^{-Ar^2}\left[B+B^2r^2+\left( B-A \right)-ABr^2\right]. \label{fj}
\end{equation}

We impose center and boundary conditions
on $E(r)$ and $p_r(r)$ respectively:
\begin{equation}
E(0)=0, \label{cc}
\end{equation}
\begin{equation}
p_r(a)=0,   \label{bc}
\end{equation}
where $a$ is a positive constant and $r=a$ defines
the charged fluid-vacuum interface.

From (\ref{fe1})-(\ref{fe3}), (\ref{ff})-(\ref{fj}), (\ref{cc}) and  (\ref{bc}) we obtain
\begin{equation}
8\pi\rho(0)=3A, \label{rc}
\end{equation}
\begin{equation}
4\pi \rho(a)=\left(A+B\right) e^{-a^2A}, \label{ra}
\end{equation}
\begin{equation}
8\pi p_r(0)=2B-A, \label{prc}
\end{equation}
\begin{equation}
8\pi p_t(0)=2B-A, \label{ptc}
\end{equation}
\begin{equation}
8\pi p_t(a)=e^{-a^2A}\left[4B-A+a^2\left(B^2-AB\right)+\frac{1}{a^2} \right]-\frac{1}{a^2}, \label{pta}
\end{equation}
\begin{equation}
E^2(a)=\frac{1}{a^2}- e^{-a^2A}\left(2B+\frac{1}{a^2}\right). \label{e2a}
\end{equation}

From (\ref{prc}) and (\ref{ptc}) we conclude that only one
pressure value is associated with $r=0$. Hence $p_r(0)$ and
$p_t(0)$ denote the same quantity i.e. central pressure.

Equations (\ref{fe1})-(\ref{fe3}) yield general expressions
for $p_t$ and $E^2$ namely
\begin{equation}
    p_t=\frac{j(r)-f(r)}{8\pi}+\rho, \label{dpt}
\end{equation}
\begin{equation}
    E^2=f(r)-8\pi\rho, \label{de2}
\end{equation}
where $\rho$ is still undetermined.

\section{LINEAR OR NON-LINEAR EQUATION OF STATE}

At this point we select an EOS with the general form
\begin{equation}
p_r=p_r(\rho,\alpha_1, \alpha_2), \label{geos}
\end{equation}
where $\alpha_1$ and $\alpha_2$ are constant parameters.

These parameters are constrained by the system of equations
\begin{equation}
p_r(0)=p_r\left[\rho(0),\alpha_1, \alpha_2\right], \label{ce1}
\end{equation}
\begin{equation}
0=p_r\left[\rho(a),\alpha_1, \alpha_2\right].       \label{ce2}
\end{equation}

Combining (\ref{fe1}) and (\ref{fe2}) we get
\begin{equation}
\rho+p_r=\frac{f(r)+h(r)}{8\pi}, \label{rpfh}
\end{equation}
which may be solved with the assumed EOS to generate specific
forms of $\rho$ and $p_r$. The arising $\rho$ is put into
(\ref{dpt}) and (\ref{de2}) to yield general expressions for $p_t$
and $E^2$.

Firstly we consider the linear EOS
\begin{equation}
p_r=\alpha_1+\alpha_2\rho. \label{leos}
\end{equation}
Following the outlined procedure we obtain
\begin{equation}
\rho=\frac{\frac{1}{8\pi}\left[f(r)+h(r)\right]-\alpha_1}{1+\alpha_2}, \label{rl}
\end{equation}
\begin{equation}
p_r=\frac{\alpha_1+\frac{\alpha_2}{8\pi}\left[f(r)+h(r)\right]}{1+\alpha_2}, \label{prl}
\end{equation}
\begin{equation}
p_t=\frac{j(r)+h(r)+\alpha_2\left[j(r)-f(r)\right]-8\pi\alpha_1}{8\pi(1+\alpha_2)}, \label{ptl}
\end{equation}
\begin{equation}
E^2=\frac{8\pi\alpha_1+\alpha_2 f(r)-h(r)}{1+\alpha_2}. \label{e2l}
\end{equation}
Taking into account (\ref{leos}) we solve the system (\ref{ce1})-(\ref{ce2})  and get
\begin{eqnarray}
\alpha_1=-\frac{\rho(a)p_r(0)}{\rho(0)-\rho(a)}, & & \alpha_2=\frac{p_r(0)}{\rho(0)-\rho(a)}. \label{a1a2}
\end{eqnarray}

Secondly we deal with the nonlinear EOS
\begin{equation}
p_r=\beta_1+\frac{\beta_2}{\rho^n}, \label{nleos1}
\end{equation}
where $n\neq-1$. Combining (\ref{nleos1}) with (\ref{rpfh}) we
obtain a polynomial equation for $\rho$ which is quadratic only
for $n=-2$ or $n=1$. The second choice for $n$ yields
\begin{equation}
\rho=\frac{k(r)-\beta_1\pm\sqrt{\left[k(r)-\beta_1\right]^2-4\beta_2}}{2}, \label{rnleos1}
\end{equation}
where
\begin{equation}
 k(r)=\frac{f(r)+h(r)}{8\pi}. \label{kdef}
\end{equation}
Solving (\ref{ce1}) and (\ref{ce2}) with (\ref{nleos1}) and $n=1$ we get
\begin{eqnarray}
\beta_1=\frac{\rho(0)p_r(0)}{\rho(0)-\rho(a)}, & & \beta_2=-\frac{\rho(0)\rho(a)p_r(0)}{\rho(0)-\rho(a)}.
     \label{b1b2}
\end{eqnarray}
If $\beta_2<0$ then each root in (\ref{rnleos1}) has definite sign.
Particularly, the root with the positive radical
determines a positive definite matter density.

Equation (\ref{nleos1}) is a modification of the Chaplygin gas EOS
used by Bertolami and P{\'a}ramos to describe neutral dark stars
\cite{bp1}. Actually, the Chaplygin gas EOS is polytropic with
negative polytropic index. The additional term $\beta_1$ allows
for a charged fluid-vacuum interface where $p_r=0$.

Thirdly we consider the modified Chaplygin EOS
\begin{equation}
p_r=\gamma_1\rho+\frac{\gamma_2}{\rho} \label{nleos2}
\end{equation}
used in the study of static, neutral, phantom-like sources
presented by Jamil, Farooq and Rashid \cite{jfr}.

Equations (\ref{ce1}) and (\ref{ce2}) combined with (\ref{nleos2})
yield
\begin{eqnarray}
\gamma_1=\frac{\rho(0)p_r(0)}{\rho(0)^2-\rho(a)^2}, & & \gamma_2=-\frac{\rho(0)p_r(0)\rho(a)^2}{\rho(0)^2
-\rho(a)^2}. \label{g1g2}
\end{eqnarray}
Putting (\ref{nleos2}) into (\ref{rpfh}) we get a quadratic
equation for $\rho$ which admits the solutions
\begin{equation}
\rho=\frac{k(r)\pm\sqrt{k(r)^2-4\left(1+\gamma_1\right)\gamma_2}}{2\left(1+\gamma_1\right)}. \label{rnleos2}
\end{equation}
If $1+\gamma_1>0$ and $\gamma_2<0$ then the sign of the radical in
(\ref{rnleos2}) uniquely determines the sign of $\rho$. As in the
previous case, a positive radical leads to positive definite
matter density.

Finally, the expressions for $\rho$ selected from (\ref{rnleos1})
and (\ref{rnleos2}) are combined with the corresponding EOS as
well as (\ref{dpt}) and (\ref{de2}) to yield solutions for $p_r$,
$p_t$ and $E^2$.

It is clear that the constant parameters appearing in (\ref{geos})
play a central role in our extension of the KB method. Their
values determine the existence of positive and negative definite
matter densities $\rho(r)$. Furthermore they lead to numerical
values for quantities like $\frac{d p_r}{d \rho}$, which is
associated with sound propagation in anisotropic charged fluids
\cite{pdl}.

\section{JUNCTION CONDITIONS AND ADIMENSIONAL FORMULATION}

Equations (\ref{a1a2}), (\ref{b1b2}), (\ref{g1g2}) show that the
constant parameters are determined by $\rho(0)$,
$\rho(a)$ and $p_r(0)$, which, according to (\ref{rc}) and
(\ref{prc}), are functions of the KB constants $A$ and $B$
appearing in (\ref{kb1}) and (\ref{kb2}). Notably, these two
constants as well as $C$ are fixed by suitable junction conditions
imposed on the internal and external metrics at $r=a$.

The external Reissner-Nordstr{\"o}m (RN) metric is given by
(\ref{le}) with
\begin{eqnarray}
e^{\nu(r)}=1-\frac{2m}{r}+\frac{q^2}{r^2}, & & e^{\lambda(r)}=\left[ 1-\frac{2m}{r}+\frac{q^2}{r^2} \right]^{-1}.
 \label{rn}
\end{eqnarray}
Junevicus \cite{jnvc} derived expressions for $A, B, C$ from the
continuity of the first and second fundamental forms across the
surface of the charged fluid sphere. In terms of the dimensionless
parameters $\mu=\frac{m}{a}$ and $\chi=\frac{|q|}{a}$ his results
take the form
\begin{equation}
A=-\frac{1}{a^2}\ln\left(1-2\mu+\chi^2\right), \label{aj}
\end{equation}
\begin{equation}
B=\frac{1}{a^2}\left(\frac{\mu-\chi^2}{1-2\mu+\chi^2}\right), \label{bj}
\end{equation}
\begin{equation}
C=\frac{\chi^2-\mu}{1-2\mu+\chi^2}+\ln\left(1-2\mu+\chi^2\right).   \label{cj}
\end{equation}

Combining (\ref{e2a}) with (\ref{aj}) and (\ref{bj}) we obtain the
electrostatic energy density at the surface of the sphere namely
\begin{equation}
E(a)^2=\frac{q^2}{a^4}.     \label{e2aj}
\end{equation}
On the other hand, using (\ref{rn}) in the electrovac $(\rho=0)$
case of (\ref{em1}) we see that the same quantity evaluated in the
external ($r>0$) spacetime region is
\begin{equation}
E(r)^2=\frac{q^2}{r^4}.         \label{e2ev}
\end{equation}
These results are compatible with the continuity of the electric
field at $r=a$.

For simplicity in numerical calculations, we reformulate our
results in terms of dimensionless quantities. From (\ref{kb1}) and
(\ref{kb2}) we see that $A$, $B$ have dimension of $length^{-2}$
whereas $C$ is dimensionless. Clearly $f(r)$, $h(r)$, $j(r)$ as
well as $\rho$, $p_r$, $p_t$, $E^2$, $\sigma$ also have dimension
of $length^{-2}$. We get the dimensionless version of any variable
or constant parameter multiplying by the appropriate power of $a$.
Here adimensionality is denoted by tildes, though quantities which
are originally dimensionless are denoted by the original symbol.
The dimensionless radial marker $x=\frac{r}{a}$ is used so the
interior of the fluid sphere is described with $x\in[0,1)$.

For the first model we get
\begin{equation}
\tilde{p}_r=\tilde{\alpha}_1+\alpha_2\tilde{\rho}, \label{aleos1}
\end{equation}
\begin{equation}
\tilde{\rho}=\frac{\frac{1}{8\pi}\left[\tilde{f}(x)+\tilde{h}(x)\right]-\tilde{\alpha}_1}{1+\alpha_2}, \label{arl}
\end{equation}
\begin{equation}
\tilde{p}_r=\frac{\tilde{\alpha}_1+\frac{\alpha_2}{8\pi}\left[\tilde{f}(x)+\tilde{h}(x)\right]}{1+\alpha_2},
 \label{aprl}
\end{equation}
\begin{equation}
\tilde{p}_t=\frac{\tilde{\j}(x)+\tilde{h}(x)+\alpha_2\left[\tilde{\j}(x)-\tilde{f}(x)\right]-
8\pi\tilde{\alpha}_1}{8\pi(1+\alpha_2)}, \label{aptl}
\end{equation}
\begin{equation}
\tilde{E}^2=\frac{8\pi\tilde{\alpha}_1+\alpha_2 \tilde{f}(x)-\tilde{h}(x)}{1+\alpha_2}, \label{ae2l}
\end{equation}
\begin{eqnarray}
\tilde{\alpha}_1=-\frac{\tilde{\rho}(1)\tilde{p}_r(0)}{\tilde{\rho}(0)-\tilde{\rho}(1)}, & &
 \alpha_2=\frac{\tilde{p}_r(0)}{\tilde{\rho}(0)-\tilde{\rho}(1)}, \label{aa1a2}
\end{eqnarray}
where
\begin{equation}
    \tilde{f}(x)=e^{-\tilde{A}x^2}\left(2\tilde{A}-\frac{1}{x^2}\right)+\frac{1}{x^2}, \label{aff}
\end{equation}
\begin{equation}
    \tilde{h}(x)=e^{-\tilde{A}x^2}\left(2\tilde{B}+\frac{1}{x^2}\right)-\frac{1}{x^2}, \label{afh}
\end{equation}
\begin{equation}
    \tilde{\j}(x)=\frac{e^{-\tilde{A}x^2}}{2}\left[2\tilde{B}+2\tilde{B}^2x^2+2\left( \tilde{B}-
    \tilde{A} \right)-2\tilde{A}\tilde{B}x^2\right], \label{afj}
\end{equation}
\begin{equation}
\tilde{\rho}(0)=\frac{3\tilde{A}}{8\pi}, \label{arc}
\end{equation}
\begin{equation}
 \tilde{\rho}(1)=\frac{\left(\tilde{A}+\tilde{B}\right) e^{-\tilde{A}}}{4\pi}, \label{ar1}
\end{equation}
\begin{equation}
 \tilde{p}_r(0)=\frac{2\tilde{B}-\tilde{A}}{8\pi}, \label{aprc}
\end{equation}
\begin{eqnarray}
\tilde{A}=a^2A, & & \tilde{B}=a^2B. \label{atbt}
\end{eqnarray}

The second model is described with
\begin{equation}
\tilde{p}_r=\tilde{\beta}_1+\frac{\tilde{\beta}_2}{\tilde{\rho}}, \label{anleos1}
\end{equation}
\begin{equation}
\tilde{\rho}=\frac{\tilde{k}(x)-\tilde{\beta}_1\pm\sqrt{\left[\tilde{k}(x)-\tilde{\beta}_1
\right]^2-4\tilde{\beta}_2}}{2},  \label{arnleos1}
\end{equation}
\begin{eqnarray}
\tilde{\beta}_1=\frac{\tilde{\rho}(0)\tilde{p}_r(0)}{\tilde{\rho}(0)-\tilde{\rho}(1)}, & &
\tilde{\beta}_2=-\frac{\tilde{\rho}(0)
\tilde{\rho}(1)\tilde{p}_r(0)}{\tilde{\rho}(0)-\tilde{\rho}(1)}. \label{ab1b2}
\end{eqnarray}

For the third model we obtain
\begin{equation}
\tilde{p}_r=\gamma_1\tilde{\rho}+\frac{\tilde{\gamma}_2}{\tilde{\rho}}, \label{anleos2}
\end{equation}
\begin{equation}
\tilde{\rho}=\frac{\tilde{k}(x)\pm\sqrt{\tilde{k}(x)^2-4\left(1+\gamma_1\right)\tilde{\gamma}_2}}
{2\left(1+\gamma_1\right)}, \label{arnleos2}
\end{equation}
\begin{eqnarray}
\gamma_1=\frac{\tilde{\rho}(0)\tilde{p}_r(0)}{\tilde{\rho}(0)^2-\tilde{\rho}(1)^2}, & & \tilde{\gamma}_2=
-\frac{\tilde{\rho}(0)\tilde{p}_r(0)\tilde{\rho}(1)^2}{\tilde{\rho}(0)^2-\tilde{\rho}(1)^2}. \label{ag1g2}
\end{eqnarray}

For models with nonlinear EOS, $\tilde{p}_t$ and $\tilde{E}^2$ are
calculated with the dimensionless versions of (\ref{dpt}) and
(\ref{de2}) respectively.

From (\ref{e2aj}) we get
\begin{equation}
\tilde{E}(1)^2=\chi^2    \label{aechi}
\end{equation}
which applies to the three models.

Finally the dimensionless proper charge density is given by
\begin{equation}
\tilde{\sigma}=\frac{e^{-\frac{\tilde{A}x^2}{2}}}{4\pi x^2}\frac{d}{dx}\left( x^2 \tilde{E} \right).        \label{aem4}
\end{equation}

\section{ANALYSIS OF THREE MODELS}

We consider only external solutions (\ref{rn}) which exclude
horizons. The standard analysis of the roots of $g_{00}=0$ implies
that the values of $\chi$ are restricted by the values of $\mu$.
Three cases arise:
\begin{enumerate}
    \item  $\mu\in(0,1)$, ~ $2\mu-1<\chi^2<\mu^2$;
    \item  $\mu\in(0,1)$, ~  $\chi=\mu$;
    \item  $\mu>0$, ~ $\chi>\mu$.
\end{enumerate}

From equations (\ref{rc}) and (\ref{aj}) we see that $\rho(0)$ is
positive if and only if $2\mu-1<\chi^2<2\mu$, which we admit as an
additional restriction on $\chi$.

The geometric mass in (\ref{rn}) is given by $m=\frac{GM}{c^2}$,
where $G$ is the gravitational constant, $M$ is the mass of the
charged source, and $c$ is the speed of light - all in
conventional units. If the mass $M$ of the charged star equals one
solar mass and $a=10$ Km, then $\mu=\frac{GM}{c^2a}\approx 0.147$.
Hence $2\mu-1$ is negative and $\sqrt{2\mu}\approx 0.543$.

Using (\ref{arc}) and (\ref{aprc}) the central density
$\tilde{\rho}(0)$ and central pressure $\tilde{p}_r(0)$ can be
evaluated for $\mu=0.147$ and $\chi\in[0,0.543)$. The monotonic
decreases of these two parameters with increasing $\chi$ are shown
in Figures 1 and 2. We observe that $\tilde{p}_r(0)$ is positive
only when $\chi$ takes values in the range $0\leq\chi<\chi_0$,
where $\chi=\chi_0$ is the only zero of $\tilde{p}_r (0)$ in the
interval $\chi\in[0,0.543)$. It is approximately given by
$\chi_0\approx0.191$.

For fixed $\mu$ we see that $\tilde{\rho}(0)$, $\tilde{\rho}(1)$
and $\tilde{p}_r(0)$ depend only on $\chi$. Hence equations
(\ref{aa1a2}), (\ref{ab1b2}) and (\ref{ag1g2}) provide expressions
for the EOS parameters as functions of $\chi$. As shown in Figures
3-5, $\tilde{\beta}_2(\chi)<0$, $\gamma_1(\chi)>0$ and
$\tilde{\gamma}_2 (\chi)<0$ for $\mu=0.147$ and
$\chi\in[0,\chi_0)$. These results allow us to select the positive
definite roots  from (\ref{rnleos1}) and (\ref{rnleos2}), and
discuss models with $\tilde{\rho}(x)>0$.

For concreteness we carry out the numerical and graphical analysis
of models with $\mu=0.147$ and $\chi\in[0,\chi_0)$, characterized
by positive matter density and positive pressure at $x=0$. Clearly
$\chi=0$ describes sources with zero net charge. This condition is
compatible with non-zero proper charge  densities
$\tilde{\sigma}(x)$. Higher values of $\chi$ are associated with
increasingly repulsive electrostatic forces that affect pressure
and matter density profiles. We aim to find out how the different
forces which allow equilibrium configurations accommodate to
varying net charge.

To begin with we examine sources with linear EOS and a selection
of $\chi$ values. The corresponding profiles for matter density,
radial and tangential pressures, and electric field are displayed
in Figures 6-9. We observe decreasing values of matter density and
pressure associated with increasing values of $\chi$. Figure 9
shows electric field profiles satisfying $\tilde{E}(0)=0$ and
$\tilde{E}(1)=\chi$. Very large negative gradients of $\tilde{E}$
develop near the surface of the sphere for $\chi=0$, and
$\tilde{E}$ increases at every $x\in(0,1]$ with increasing $\chi$.
Notably Figure 7 indicates that increasing net charges and
electric fields are associated with decreasing radial pressure
gradients at each point.

Figures 10 and 11 show profiles for the squares of radial and
tangential sound speeds, defined by $v_{sr}^2=\frac{dp_r}{d\rho}$
and $v_{st}^2=\frac{dp_t}{d\rho}$ respectively. We observe that
$v_{sr}^2$ is independent of $x$ and decreases with increasing
$\chi$. For fixed $\chi$, $v_{st}^2$  monotonically decreases with
increasing $x$; and increases with increasing $\chi$ for fixed
$x$. These parameters satisfy the inequalities $0\leq v_{sr}^2
\leq 1$ and $0\leq v_{st}^2 \leq 1$ everywhere within the charged
fluid for the five values of $\chi$ considered.

Based on the standard analysis of energy conditions for charged
anisotropic fluids (see, for example, \cite{pdl}), we find that
the weak energy condition (WEC), the strong energy condition (SEC)
and the dominant energy condition (DEC) are simultaneously
satisfied if and only if the following six inequalities hold at
every point within the source:
\begin{equation}
    \tilde{\rho}+\tilde{p}_r\geq 0, \label{ec1}
\end{equation}
\begin{equation}
    \tilde{\rho}+\frac{\tilde{E}^2}{8\pi}\geq 0, \label{ec2}
\end{equation}
\begin{equation}
    \tilde{\rho}+\tilde{p}_t+\frac{\tilde{E}^2}{4\pi}\geq 0, \label{ec3}
\end{equation}
\begin{equation}
    \tilde{\rho}+\tilde{p}_r+2\tilde{p}_t+\frac{\tilde{E}^2}{4\pi}\geq 0,   \label{ec4}
\end{equation}
\begin{equation}
\tilde{\rho}+\frac{\tilde{E}^2}{8\pi}-|\tilde{p}_r-\frac{\tilde{E}^2}{8\pi}|\geq 0, \label{ec5}
\end{equation}
\begin{equation}
\tilde{\rho}+\frac{\tilde{E}^2}{8\pi}-|\tilde{p}_t+\frac{\tilde{E}^2}{8\pi}|\geq 0. \label{ec6}
\end{equation}

Inequalities (\ref{ec1}) and (\ref{ec2}) hold automatically for
the sources considered here. Straightforward plotting of
the left sides of (\ref{ec3})-(\ref{ec6}) shows
that these inequalities are satisfied as well at every
$x\in[0,1]$.

Further analysis shows that models with nonlinear EOS
(\ref{anleos1}) and (\ref{anleos2}) are very similar to those
satisfying (\ref{aleos1}). Particularly, matter densities and
radial pressures are everywhere positive, matter densities as well
as radial and tangential pressures monotonically decrease with
increasing $x$, radial and tangential sound speeds satisfy $0\leq
v_{sr}^2 \leq 1$ and $0\leq v_{st}^2 \leq 1$, $v_{sr}$ decreases
with increasing $\chi$ for fixed $x$, and WEC, SEC and DEC are
satisfied for $\mu=0.147$ and $\chi=0,0.05,0.1,0.15,0.19$. However
models with nonlinear EOS get $v_{sr}$ profiles which increase
monotonically with increasing $x$ and fixed $\chi$.

Another difference among the three models concerns the anisotropy
parameter $\tilde{\Delta}=\tilde{p}_t -\tilde{p}_r$. As shown in
Figures 12 - 14, each EOS affect the dependence of
$\tilde{\Delta}$ on $x$ in a different way. Particularly, the sign
of $\tilde{\Delta}$ is the same at every $x\in(0,1]$  for fixed
$\chi$ only in models with linear EOS. The effect of
(\ref{anleos1}) on pressure anisotropy  is notorious as three
$\tilde{\Delta}(x)$ profiles develop sign changes in that case.
However the  predominance of radial pressure over tangential
pressure for the highest value of $\chi$ is a common feature of
the three models.

Are these three types of models physically meaningful? Delgaty and
Lake found that only a relatively small number of proposed perfect
fluid sources for the Schwarzschild metric satisfy a set of well
established conditions for physical acceptability \cite{dlpa}.
These conditions include regularity of the origin, positive matter
density and pressure, decreasing matter density and pressure with
increasing $r$, causal sound propagation and smooth matching of
the internal and external metrics at the fluid-vacuum interface
$(r=a)$. Some authors add the condition of monotonically
decreasing sound speed with increasing $r$ (see, for example,
\cite{bhna}). Regarding anisotropic fluid sources, the causality
condition $0\leq v_{s}^2 \leq 1$ has been imposed on radial and
tangential sound speeds (see, for example, \cite{hmar} and
\cite{ahn}). Disagreement arises in connection with the sign of
tangential pressure which is unrestricted for some authors (see,
for example, \cite{hmar}) and strictly positive for others (see,
for example, \cite{ahn}). The above discussion indicates that our
three charged anisotropic sources satisfy most of the usual
acceptability criteria. Conflict may arise only in connection with
negative tangential pressures occurring for the highest $\chi$
values, constant sound speeds $v_{sr}$ associated with the linear
EOS, and the increase of $v_{sr}$ with increasing $x$ appearing in
the models with nonlinear EOS.

Electric interactions and charge distributions in our models
deserve further analysis. We have found that $\tilde{E}$ profiles
are affected by the choice of EOS. Particularly, in models with
$\chi=0$ the maximum $\tilde{E}$ values are approximately $0.03$,
$0.07$ and $0.05$, corresponding to (\ref{aleos1}),
(\ref{anleos1}) and (\ref{anleos2}), respectively. In the case of
EOS (\ref{aleos1}), $\tilde{\sigma}(0)$ is finite and increases
with increasing $\chi$; and $\tilde{\sigma}(1)$ is finite for
$\chi=0.05,0.1,0.15,0.19$ but unbounded for $\chi=0$. Moreover,
only sources with $\chi=0$ contain electric charge of both signs.
We could have anticipated this behaviour of $\tilde{\sigma}$ from
Maxwell equation (\ref{aem4}) and the slope changes displayed in
Figure 9. Figure 15 shows the arising $\tilde{\sigma}$ profiles in
the restricted interval $[0,0.999]$. The most important difference
among $\tilde{\sigma}$ profiles associated with the three EOS is
that models with nonlinear EOS and $\chi=0.05$ also involve
positive and negative electric charges.

We have noticed that the charge distributions considered by
Thirukkanesh and Maharaj \cite{tm1} are singular at the origin,
where the electric field does not vanish. Alternatively, all our
sources have vanishing electric field and finite proper charge
density at $r=0$. On the other hand, our models with $\chi=0$
involve charge distributions which are singular at the
fluid-vacuum interface. We remark that charged thin shells are not
considered in our study and electric fields are continuous at
$r=a$. Also, charged sources with $\chi=0$ are characterized by
$E(a)=0$ and unbounded $\lim_{r\to a^{-}} E^{\prime}(r)$.
Arbitrarily large electric field gradients near the fluid-vacuum
are puzzling and we proceed with a preliminary discussion of their
significance.

The proper charge density $\sigma$ appears explicitly in the
inhomogeneous Maxwell equation (\ref{em4}). It determines the net
charge inside a sphere of radius $r$ through the formula
\begin{equation}
q(r)=4\pi\int_{0}^{r}\sigma(r) e^{\frac{\lambda(r)}{2}} r^2 dr. \label{nq}
\end{equation}
We remark that $q(a)$ is the total charge of the source, denoted by $q$ in (\ref{rn}).
It is $q(r)$ the quantity that actually determines the electric field
\begin{equation}
E(r)=\frac{q(r)}{r^2}.  \label{efield}
\end{equation}

Proper charge density  $\sigma(r)$ is not an observable quantity
in standard EM theory. We have seen that $E(r)$ takes part in the
formulation of the energy conditions of general relativity through
the contribution of the Maxwell field to the energy-momentum
tensor. In contrast, the authors are not aware of any condition
for physical acceptability imposed directly on $\sigma$.

Both spacetime curvature and electric field gradients deviate the
worldlines of charged test particles. Balkin, van Holten and
Kerner \cite{bvhk} and van Holten \cite{vh} have derived covariant
formulae for the relative acceleration of particles with the same
charge/mass ratio. Their equations include terms containing
derivatives of the Maxwell tensor. The question arises as to whether
these terms could lead to infinite relative accelerations for
pairs of charged test particles passing through the fluid-vacuum
interface of models with $\chi=0$. Hence very large relative
accelerations of test particles could allow indirect observation
of infinite charge density located at the vacuum-fluid interface
of our sources.

The fundamental invariant of the electromagnetic field
$I_1=F_{\mu\nu}F^{\mu\nu}$ has been used in the analysis of
genuine singularities of static solutions for the Einstein-Maxwell
equations \cite{hh}. This invariant is bounded at the fluid-vacuum
interfaces of our three types of models. The inhomogeneous Maxwell
equation implies that
$I_2=F^{\mu\nu}{}_{;\nu}{F^{\omega}{}_{\mu;\omega}}$ is
proportional to $\sigma(r)^2$. The fact that $\sigma(r)$ is
unbounded at $r=a$ in models with $\chi=0$ determines the singular
behaviour of a differential invariant of the theory. Questions
about physical acceptability of solutions with regular $I_1$ and
singular $I_2$ are analogous to those regarding spacetimes with
regular curvature invariants and singular differential invariants
discussed by Musgrave and Lake \cite{ml}.

The discussion of physical acceptability for sources with zero net
charge points at the equilibrium condition for sections with
infinite charge density. Fluid elements with unbounded $\sigma$
are located at $r=a$, where the electric field vanishes due to the
choice $\chi=0$. Nevertheless, the conclusion that no electric
force acts on elements of charged fluid with infinite $\sigma$ is
not straightforward and deserves closer examination. Other
features of our models motivate further analysis of equilibrium
conditions. For example, as we keep $\mu$ constant and increase
$\chi$, repulsive electric forces increase with decreasing matter
density, decreasing pressure and decreasing radial pressure
gradients. The question is, how gravitational and other fluid
forces counteract increasing electrostatic repulsion when the
charged fluid becomes more diluted and pressure gradients tend to
vanish? The situation is clarified using the generalized
Tolman-Oppenheimer-Volkov (TOV) equation as presented by Ponce de
Le\'{o}n \cite{pdl}. (See also \cite{gmkps}.) It reads
\begin{equation}
-\frac{M_G\left(\rho+p_r\right)}{r^2}e^{\frac{\lambda-\nu}{2}}-\frac{dp_r}{dr}+\sigma
\frac{q }{r^2}e^{\frac{\lambda}{2}}+\frac{2}{r}\left(p_t-p_r\right)=0,    \label{tov}
\end{equation}
where $M_G=M_G(r)$ is the effective gravitational mass inside a
sphere of radius $r$ and $q=q(r)$ is given by (\ref{nq}). The
effective gravitational mass is given by the expression
\begin{equation}
    M_G(r)=\frac{1}{2}r^2e^{\frac{\nu-\lambda}{2}}\nu^{\prime}, \label{egm}
\end{equation}
derived from the Tolman-Whittaker formula and the Einstein-Maxwell
equations \cite{pdl}.

Equation (\ref{tov}) expresses the equilibrium condition for
charged fluid elements subject to gravitational, hydrostatic and
electric forces, plus another force due to pressure anisotropy.
Combined with (\ref{kb1}), (\ref{kb2}), (\ref{atbt}) and
(\ref{efield}), it takes the adimensional form
\begin{equation}
\tilde{F}_g+\tilde{F}_h+\tilde{F}_e+\tilde{F}_a=0,      \label{afnt}
\end{equation}
where
\begin{equation}
\tilde{F}_g=-\tilde{B}x\left(\tilde{\rho}+\tilde{p}_r\right),   \label{afgt}
\end{equation}
\begin{equation}
\tilde{F}_h=-\frac{d\tilde{p}_r}{dx},   \label{afht}
\end{equation}
\begin{equation}
\tilde{F}_e=\tilde{\sigma}\tilde{E}e^{\frac{\tilde{A}x^2}{2}}, \label{afet}
\end{equation}
\begin{equation}
\tilde{F}_a=\frac{2}{x}\left(\tilde{p}_t-\tilde{p}_r\right).        \label{afat}
\end{equation}

The profiles of $\tilde{F}_g$, $\tilde{F}_h$, $\tilde{F}_e$ and
$\tilde{F}_a$ for sources with linear EOS, $\mu=0.147$ and
$\chi=0$ are shown in Figure 16. Notably, the electric force
acting on fluid elements with unbounded $\tilde{\sigma}$ located
at $x=1$ is finite although $\tilde{E}(1)=0$ in this case.
Equivalently $\lim_{x \to 1} \tilde{\sigma}\tilde{E}$ is finite.
Furthermore $\tilde{F}_e$ is the weakest force and changes sign at
$x\approx 0.85$. Both $\tilde{F}_h$ and $\tilde{F}_a$ point
outwards at every $x\in(0,1]$. Electric forces increase and radial
pressure gradients decrease with increasing $\chi$. When
$\chi=0.19$ the hydrostatic force $\tilde{F}_h$ is relatively
insignificant and $\tilde{F}_a$ points inwards, acting in
conjunction with gravitational attraction to compensate the
electrostatic repulsion. This situation is described in Figure 17.
Clearly, the sign of $\tilde{F}_a$ changes due to the predominance
of $p_r$ over $p_t$ for the largest $\chi$ values. This sign
inversion is essential for the configuration of our static,
charged anisotropic sources with linear EOS.

We have extended the analysis of these four forces to models with
nonlinear EOS and found essentially the same equilibrium
configuration discussed above for linear EOS and $\chi=0.19$.
Particularly, $\tilde{F}_h$ is comparatively small, $\tilde{F}_g$
and $\tilde{F}_a$ have nearly  the same profile and jointly
counteract the electrostatic repulsion. Hence the choice of EOS
has  a negligible effect on the compensation of relatively large
electrostatic repulsion by gravitational attraction and pressure
anisotropy.

Equations (\ref{em1})-(\ref{em4}) indicate that matter density,
radial pressure, tangential pressure and electric field strength
affect the spacetime metric in our relativistic fluid models. Our
choice $\mu=0.147$ corresponds to a compact stellar object
($M=2\times10^{33}$ g, $a=10^6$ cm), so pressure is expected to
play a substantial role here. The question arises as to whether
the electrostatic energy density significantly contributes to the
source of gravity. We compute dimensionless values of the physical
variables and get maxima and minima for $\tilde{\rho}_{0}$
(central matter density), $\tilde{p}_{0}$ (central pressure),
$\tilde{p}_{ta}$ (tangential pressure at $r=a$) and
$\tilde{E}_{a}^{2}$ (electrostatic energy density at $r=a$) as
well as $\tilde{E}_{m}^{2}$ (maximum electrostatic energy density
for $\chi=0$). Approximate numerical results for the model with
linear EOS are displayed in Table I.

\begin{table}
\caption{Dimensionless values of physical variables for the model
with linear EOS.}
\begin{tabular}{|c|c|c|c|c|c|}
\hline
$\chi$ & $\tilde{\rho}_{0}$ & $\tilde{p}_{0}$ & $\tilde{p}_{ta}$ & $\tilde{E}_{a}^{2}$ & $\tilde{E}_{m}^{2}$ \\
\hline
0 & 0.042 & 0.003 & 0.001 & 0 & 0.0009  \\
\hline
0.19 & 0.036 & 0.00005 & -0.002 & 0.036 & - \\
\hline
\end{tabular}
\label{Table I}
\end{table}

We observe that the maximum values of $\tilde{p}_{0}$ and
$\tilde{p}_{ta}$ are just one order of magnitude smaller than the
maximum of $\tilde{\rho}_{0}$. These maxima correspond to the
source with $\chi=0$ i.e. zero net charge and weakest
electrostatic repulsion. Notably, these density and pressure
values are very similar to the dimensionless density
$\tilde{\rho}_{s}$ and dimensionless central pressure
$\tilde{p}_{0s}$ of the (uncharged) Schwarzschild internal
solution (SIS) with the same $\mu$ value. Using well-known
formulae for the SIS in geometrized units \cite{grw} we derive the
dimensionless expressions
\begin{equation}
\tilde{\rho}_{s}=\frac{3\mu}{4\pi},
\end{equation}
\begin{equation}
\tilde{p}_{0s}=\frac{3\mu^2}{8\pi},
\end{equation}
leading to the approximate results $\tilde{\rho}_{s}=0.036$ and
$\tilde{p}_{0s}=0.003$.

The value of $\tilde{E}_{m}^{2}$ shown in Table I indicates a
substantial contribution of the Maxwell field to the source of
gravity in the case of zero net charge, although matter density
and pressure predominate.

In sources with maximum net charge ($\chi=0.19$) $\tilde{p}_{0}$
decreases in two orders of magnitude, $\tilde{p}_{ta}$ doubles its
absolute value and gets the opposite sign, and $\tilde{\rho}_{0}$
and $\tilde{E}_{a}^{2}$ have the same approximate value. From
figures 13 and 14 we see that the reduction of radial pressure
with increasing net charge makes gravitational attraction weaker
in spite of the larger values of electrostatic energy density. We
have discussed above that the sign inversion of tangential
pressure leads to an extra force which complements gravitational
attraction, so that the stronger electric repulsion gets
compensated. Certainly static equilibrium is attainable in these
charged sources thanks to pressure anisotropy.

The discussion of the above quantities in conventional units is
interesting as well. Adimensional density and pressure values are
expressed in geometrized units and then converted to cgs units
using well-known conversion factors \cite{mtw}. Similarly,
adimensional values of electrical variables are expressed in
geometrized units, then converted to Heaviside-Lorentz units and
finally to Gaussian-cgs units. The corresponding results are
displayed in tables II and III.

\begin{table}
\caption{Density and pressures for the model with linear EOS expressed in cgs units.}
\begin{tabular}{|c|c|c|c|}
\hline
$\chi$ & $\rho_{0}\,(gr/cm^3)$ & $p_{0}\,(dyn/cm^2)$ & $p_{ta}\,(dyn/cm^2)$\\
\hline
0 & $5.6\times10^{14}$ & $3.6\times10^{34}$ & $1.2\times10^{34}$ \\
\hline
0.19 & $4.8\times10^{14}$ & $6\times10^{32}$ & $-2.4\times10^{34}$\\
\hline
\end{tabular}
\label{Table II}
\end{table}

\begin{table}
\caption{Electric field strengths and net charge for the model
with linear EOS expressed in Gaussian-cgs units}
\begin{tabular}{|c|c|c|c|}
\hline
$\chi$ & $E_{a}\,(statvolt/cm)$ & $E_{m}\,(statvolt/cm)$ & $q\,(statcoul)$\\
\hline
0 &  0 & $3.7\times10^{23}$ & 0 \\
\hline
0.19 & $2.3\times10^{24}$ & $-$ & $1.9\times10^{29}$ \\
\hline
\end{tabular}
\label{Table III}
\end{table}

These values of central matter density and pressure as well as
electric field strength and net charge in conventional units are
similar to those discussed by previous authors in the context of
charged compact objects with isotropic pressure \cite{reml}. It is
understood that the huge gravitational attraction determined by
these values of central matter density and pressure compensates
repulsive electrostatic forces associated with field strengths and
net charges with orders of magnitude $10^{24}\,statvolt/cm$
($10^{26}\,V/cm$) and $10^{29}\, statcoul$ ($10^{19}\,C$)
respectively. However our study of charged anisotropic sources
points at the crucial role that forces caused by pressure
anisotropy can play in the construction of equilibrium states.
This anisotropic effect is equally important for the three models
considered here, at least for the assumed value of $\mu=0.147$.
Provided that pressure anisotropy supplies an inward-directed
force that compensates electrostatic repulsion, we expect our
sources can achieve higher charges and electric field strengths
than sources with isotropic pressures and the same $\mu$ value.
Also, our results for sources with zero net charge suggest a
possible role for pressure anisotropy and electric charge in the
structure of static sources for the external Schwarzschild metric.

\section{Concluding remarks}

The linear equation (\ref{leos}) links our analytical approach
with the numerical treatment of electrically charged strange quark
stars by Negreiros et al. \cite{ecsqs} based on the MIT Bag Model.
They also assume (\ref{cc}) and impose vanishing isotropic
pressure at the fluid-vacuum interface.

We have borrowed nonlinear EOS (\ref{nleos1}) (with $n=1$) and
(\ref{nleos2}) from current models of dark matter and dark energy.
Most applications of Chaplygin and modified Chaplygin gases are
cosmological and describe non-static, neutral gravitational fluids
with isotropic pressures. Instead our bounded sources involve
static, asymptotically flat spacetimes and charged anisotropic
fluids. Actually we have targeted the effects of these EOS on the
interior regions and fluid-vacuum interfaces of charged stars
appart from any cosmological framework. The question arises as to how
dark matter-dominated star interiors could get electric charge.
Specific charge transfer mechanisms hypothesized in studies of
charged neutron stars \cite{madoha} could be considered.
Alternatively, a fraction of dark matter could be made up by
massive particles with electric charge (CHAMPs) \cite{champs}, so
dark stars could be charged. We remark that our procedures and
results are totally independent of any charge generation
mechanism.

Lobo initiated the study of van der Waals (VDW) quintessence stars
\cite{lobo}. Apart from variations of $\rho$ and $p_r$ in the
interior of the VDW fluid, which may be caused by gravitational
instabilities, this author imposes a cut-off of the
energy-momentum tensor at $r=a$, where the internal metric matches
the external Schwarzschild solution. Again, the motivation for
introducing this new type of bounded source is cosmological.
Provided that the quintessence EOS leads to interesting
descriptions of the late universe, stellar objects arising from
fluid nucleation through density perturbations are explored.

The extended KB approach can be applied to a sphere of charged
quintessence fluid. Apart from the hypothetical link of dark
matter with CHAMPs, we cannot justify the introduction of charge
in this model. However we aim to test the applicability of the
present approach against the more complicated VDW EOS
\begin{equation}
    p_r=\frac{\delta_1\rho}{1-\delta_3\rho}-\delta_2\rho^2 \label{vdweos}
\end{equation}
which describes dark matter and dark energy as a single fluid \cite{lobo}.

Assuming that the interior metric is joined to (\ref{rn}), and
that the electric field and radial pressure vanish at $r=0$ and
$r=a$ respectively, we evaluate (\ref{vdweos}) at the center and
boundary of the charged sphere and solve two linear equations for
$\delta_1$ and $\delta_2$ as functions of $\rho(0)$, $\rho(a)$,
$p_r(0)$ and $\delta_3$. Then (\ref{vdweos}) is combined with
(\ref{rpfh}) to obtain
\begin{equation}
    \delta_2\delta_3\rho^3-(\delta_2+\delta_3)\rho^2+\left[1+\delta_1+\delta_3\,g(r)\right]\rho-g(r)=0,
    \label{cubic}
\end{equation}
where $g(r)=\frac{f(r)+h(r)}{8\pi}$.

We do not go further into the analysis of the real roots of
(\ref{cubic}) and the corresponding sources. However, we
anticipate difficulties in the selection of $\mu$, $\chi$ and
$\tilde{\delta}_3$ values leading to positive definite matter
densities.

The analysis of stability is beyond the scope of this paper.
Results presented by Andr{\'e}asson \cite{hand} regarding
stability properties of charged anisotropic spheres could shed
light on this fundamental issue. We highlight the potentially
interesting study of gravitational collapse of sources with zero
net charge, and hope our results will motivate further research on
these topics.

Finally we point out that different EOS could be matched at
specific radii to provide composite models. For example, a linear
interior (core) with a nonlinear exterior (layer) matched at a
location inside the charged fluid sphere might provide a better
model than a simple EOS throughout. Using the extended KB approach
developed here, both the core and the layer would be described
with the same general metric (\ref{le}) and the {\it ans{\"a}tze}
(\ref{kb1}) and (\ref{kb2}) with presumably different values of
$A$, $B$ and $C$ in each region. Less restricted choices of these
constants could lead to charged thin shells emerging at the
interface between the fluids.

\section*{Acknowledgements}
VV is grateful to Professor Graham Hall and the Institute of
Mathematics of Aberdeen University for the Honorary Research
Fellowship granted to him since February 2006. Also he is indebted
to Dr. Olga Savasta for valuable computing assistance. FR, SR, KC
and MK are grateful to the authorities of IUCAA for research
facilities and hospitality.

\begin{figure*}[ptbh]
\includegraphics[scale=.4]{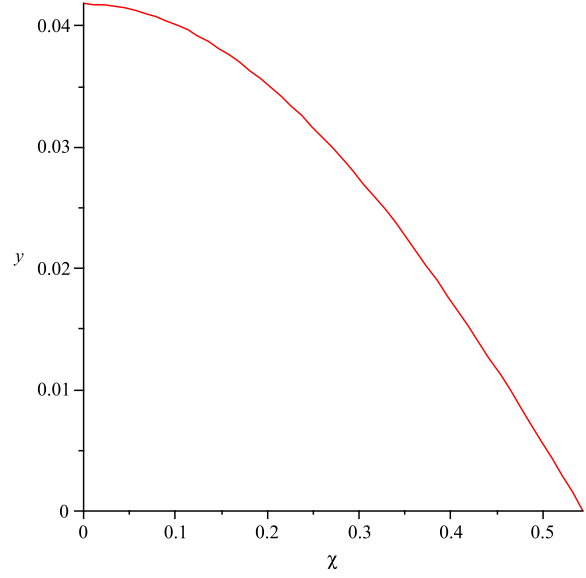}
\caption{Central density $\tilde{\rho}(0)$ as a function of $\chi$ for $\mu = 0.147$ ($y \equiv \tilde{\rho}(0)$).}
\label{Figure 1}
\end{figure*}

\begin{figure*}[ptbh]
\includegraphics[scale=.4]{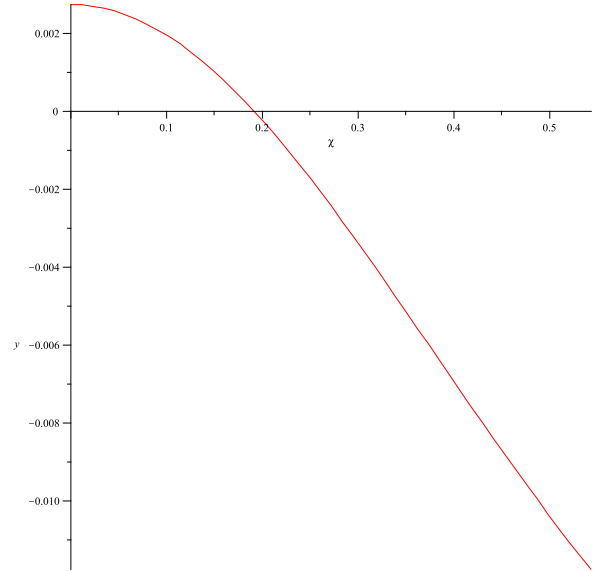}
\caption{Central pressure $\tilde{p}_r(0)$ as a function of $\chi$ for $\mu = 0.147$ ($y \equiv \tilde{p}_r(0)$).}
\end{figure*}

\begin{figure*}[ptbh]
\includegraphics[scale=.3]{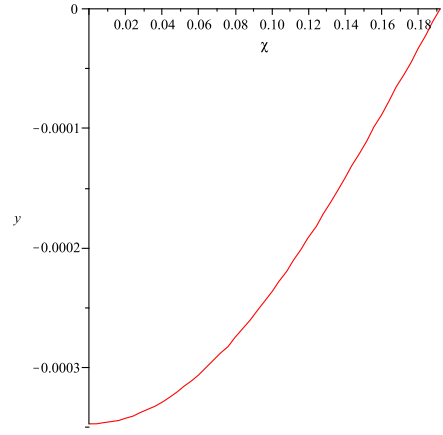}
\caption{Coefficient $\tilde{\beta_2}$ as a function of $\chi$ for
$\mu = 0.147$ ($y \equiv \tilde{\beta}_2$).}
\end{figure*}

\begin{figure*}[ptbh]
\includegraphics[scale=.3]{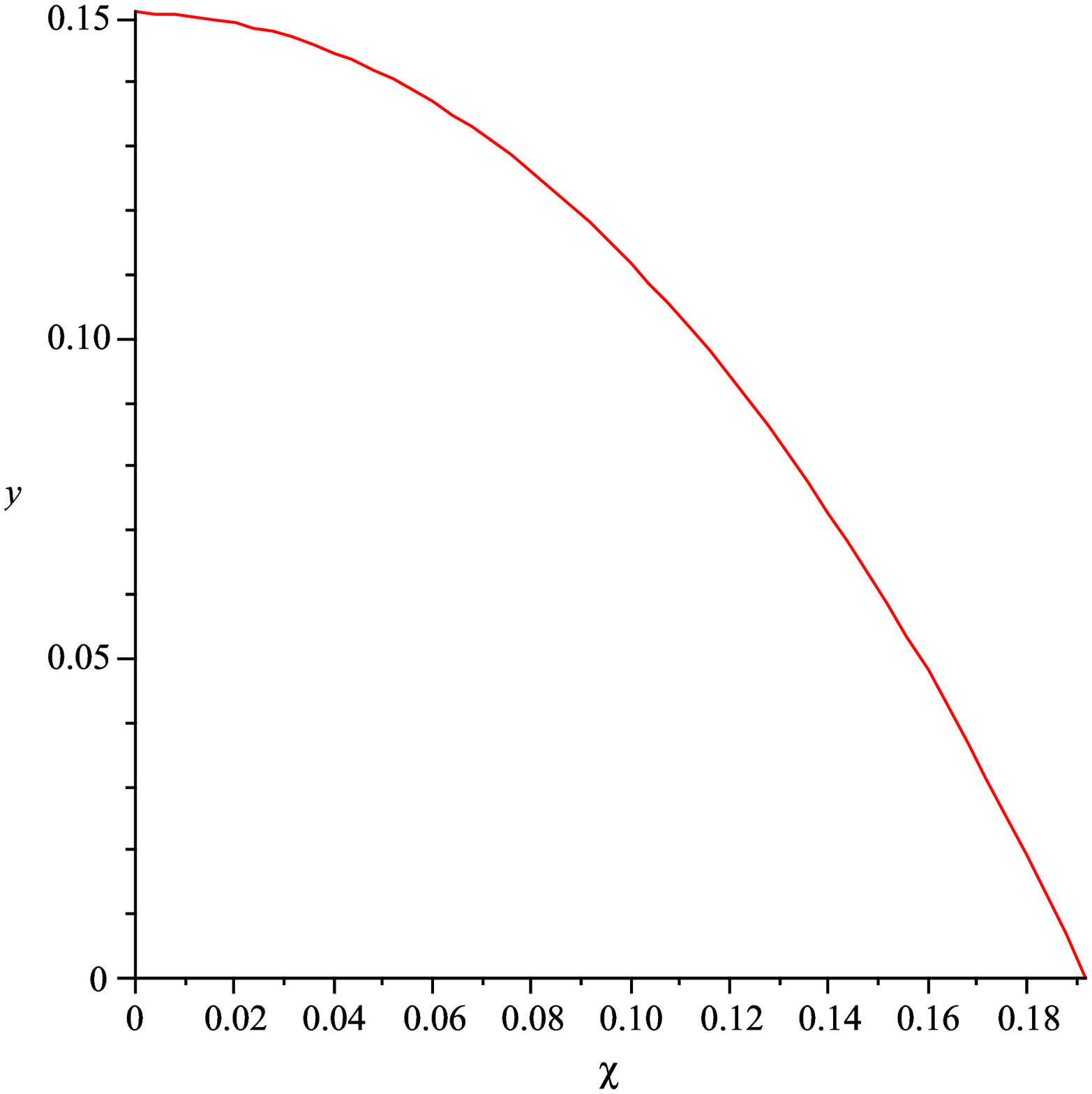}
\caption{Coefficient $\tilde{\gamma}_1$ as a function of $\chi$
for $\mu = 0.147$ ($y \equiv \tilde{\gamma}_1$).}
\end{figure*}

\begin{figure*}[ptbh]
\includegraphics[scale=.5]{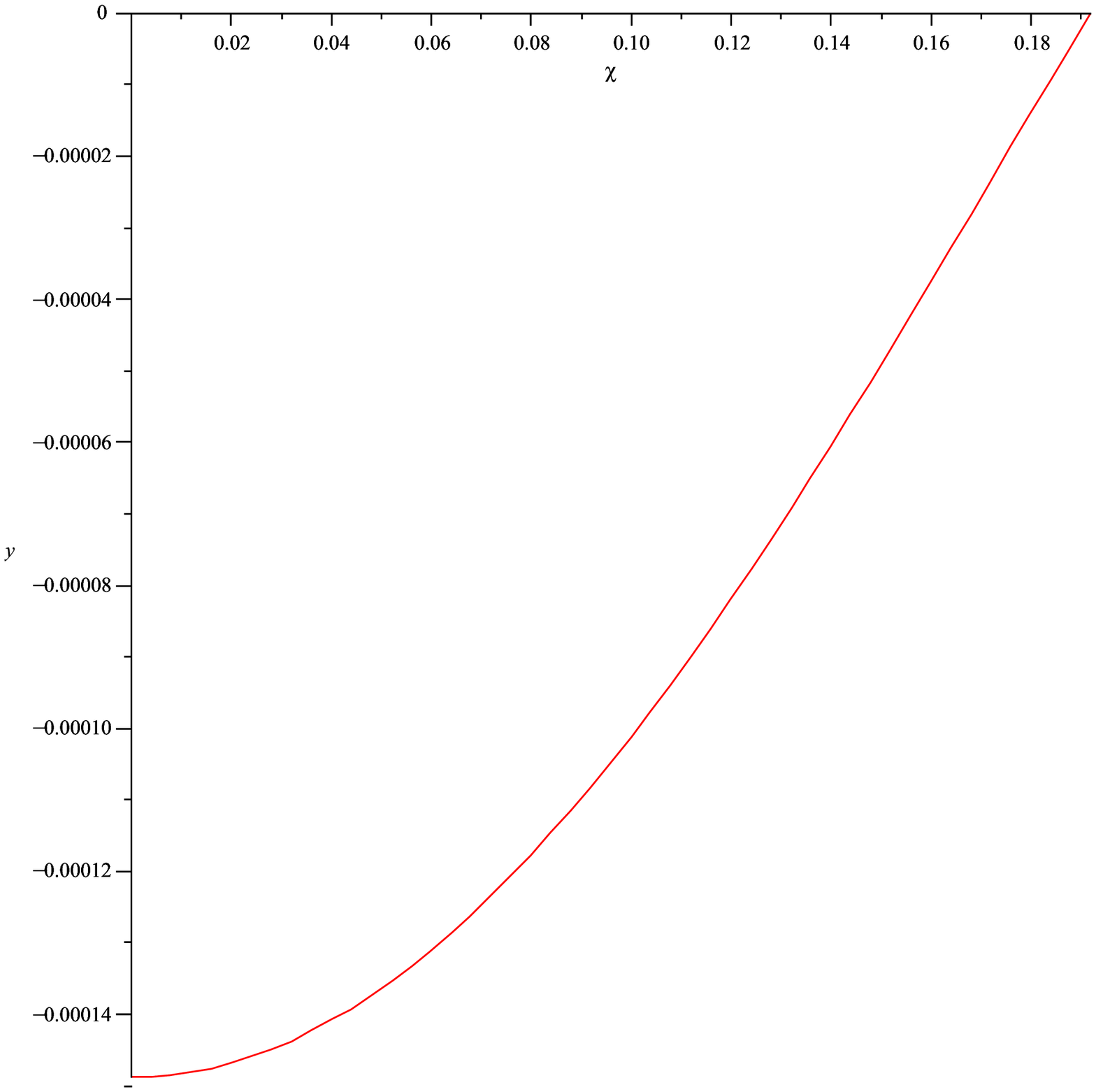}
\caption{Coefficient $\tilde{\gamma}_2$ as a function of $\chi$
for $\mu = 0.147$ ($y \equiv \tilde{\gamma}_2$).}
\end{figure*}

\begin{figure*}[ptbh]
\includegraphics[scale=.3]{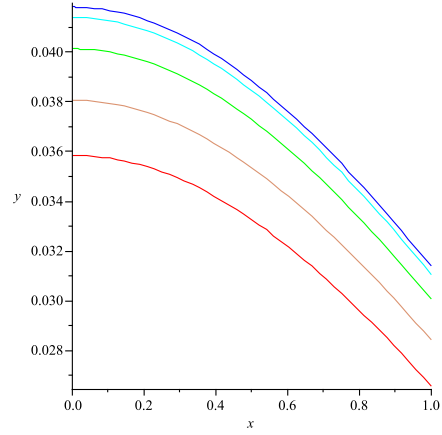}
\caption{Matter density $\tilde{\rho}$ as a function of $x$ for
five different values of $\chi$. In this figure the curves $1-5$
from the top corresponds to $\chi=0, 0.05, 0.10, 0.15$ and $0.19$
respectively ($y \equiv \tilde{\rho}$).}
\end{figure*}

\begin{figure*}[ptbh]
\includegraphics[scale=.4]{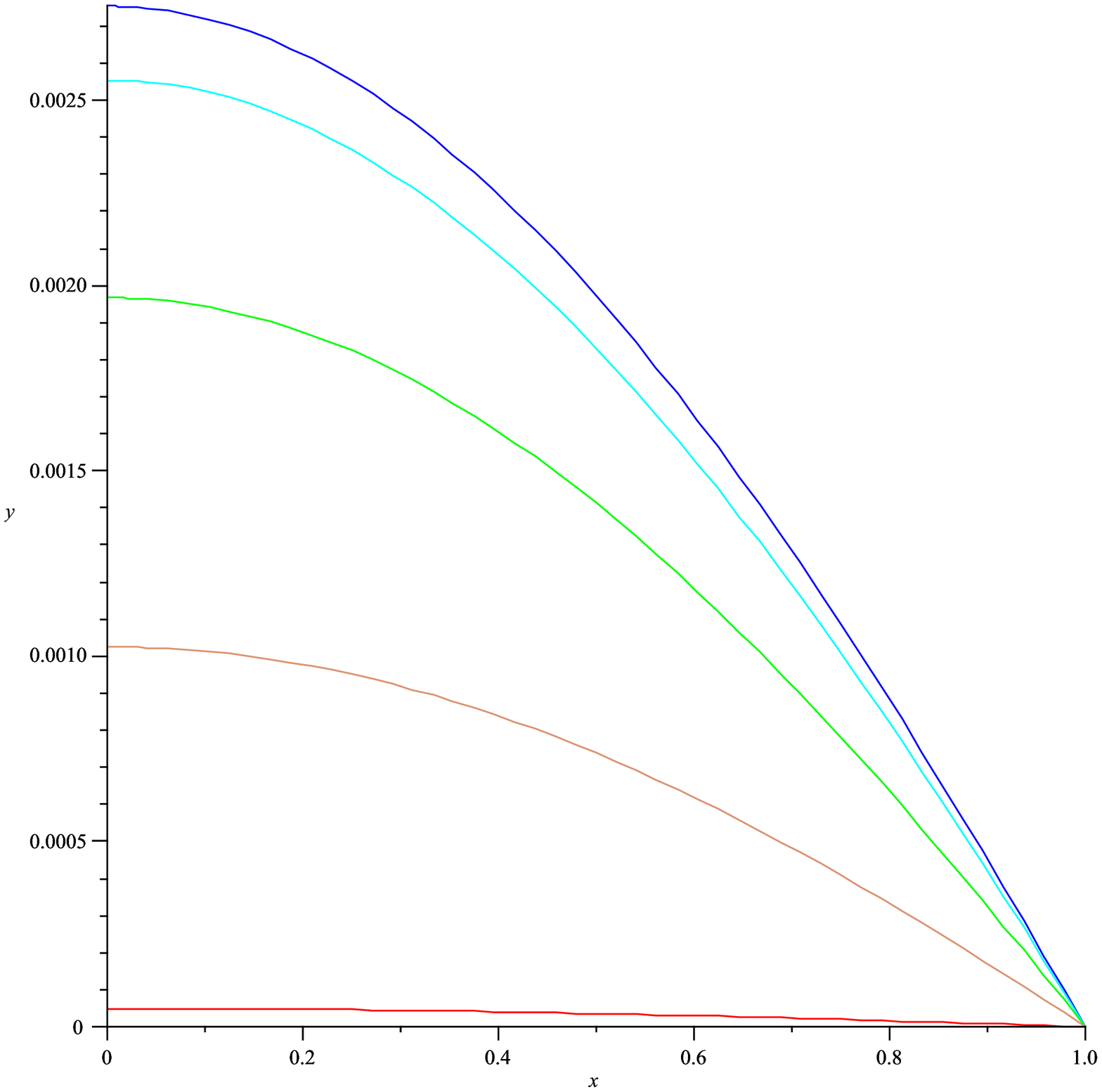}
\caption{Radial pressure $\tilde{p}_r$ as a function of $x$ for
five different values of $\chi$. In this figure the curves $1-5$
from the top correspond to $\chi=0, 0.05, 0.10, 0.15$ and $0.19$
respectively ($y \equiv \tilde{p}_r$).}
\end{figure*}

\begin{figure*}[ptbh]
\includegraphics[scale=.5]{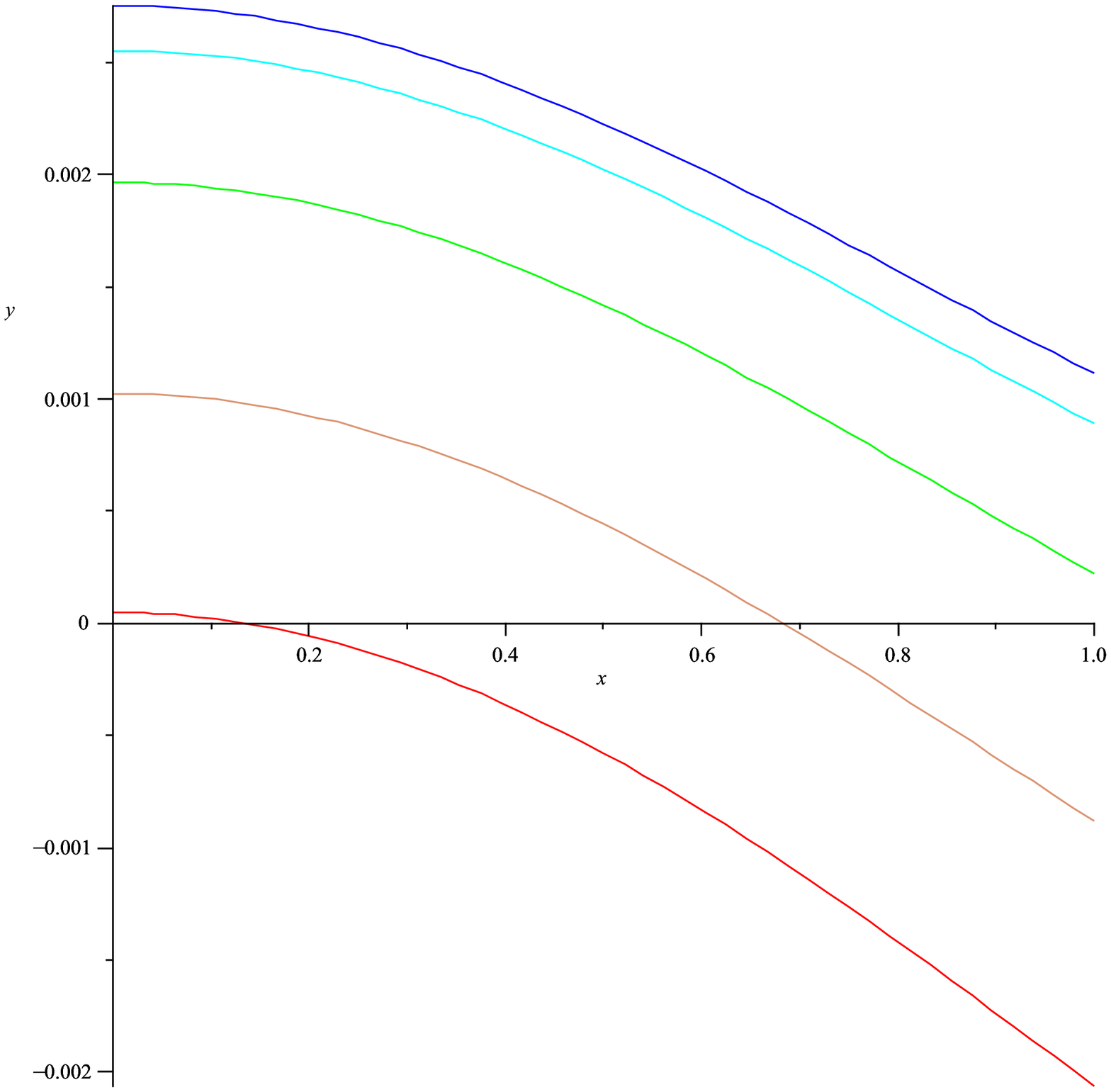}
\caption{Tangential pressure $\tilde{p}_t$ as a function of $x$
for five different values of $\chi$. In this figure the curves
$1-5$ from the top correspond to $\chi=0, 0.05, 0.10, 0.15$ and
$0.19$ respectively ($y \equiv \tilde{p}_t$).}
\end{figure*}

\begin{figure*}[ptbh]
\includegraphics[scale=.5]{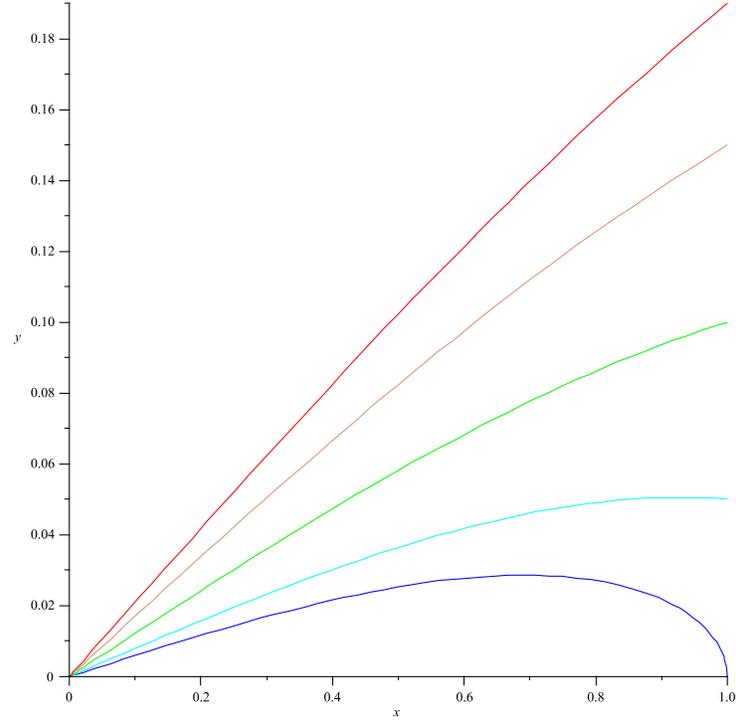}
\caption{Electric field $\tilde{E}$ as a function of $x$ for five
different values of $\chi$. In this figure the curves $1-5$ from
the bottom correspond to $\chi=0, 0.05, 0.10, 0.15$ and $0.19$
respectively ($y \equiv \tilde{E}$).}
\end{figure*}

\begin{figure*}[ptbh]
\includegraphics[scale=.5]{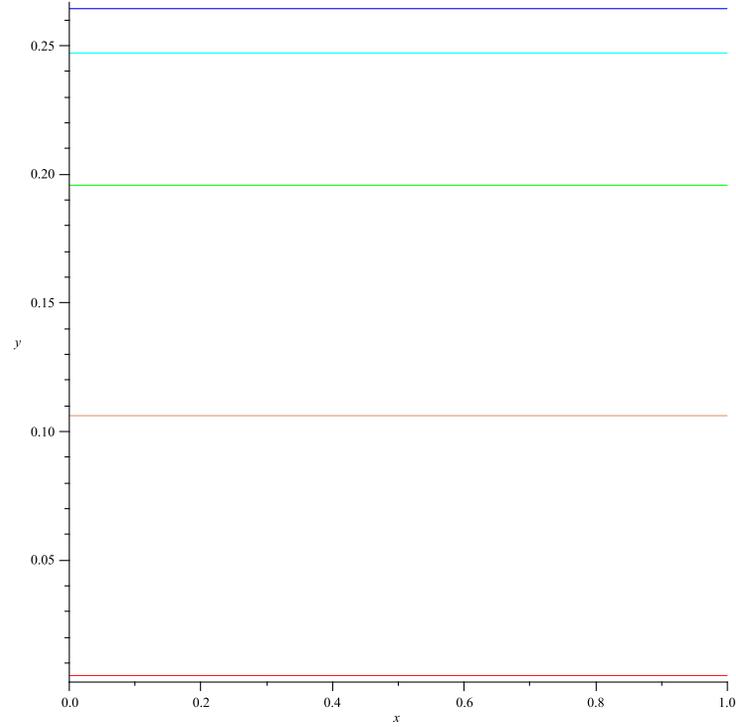}
\caption{Square of radial sound velocity $v_{sr}^2$ as a function
of $x$ for five different values of $\chi$. In this figure the
curves $1-5$ from the top correspond to $\chi=0, 0.05, 0.10, 0.15$
and $0.19$ respectively ($y \equiv v_{sr}^2$).}
\end{figure*}

\begin{figure*}[ptbh]
\includegraphics[scale=.5]{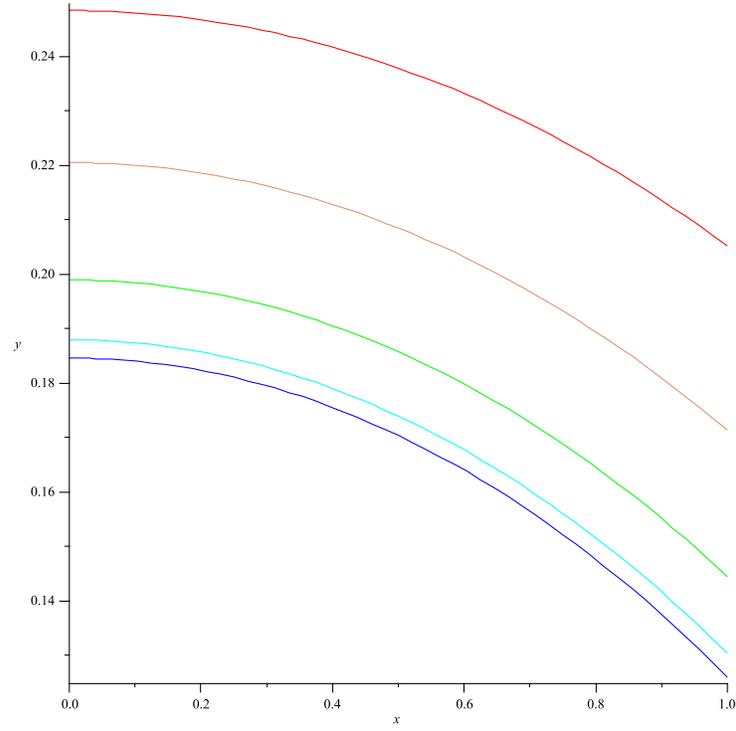}
\caption{Tangential sound velocity $v_{st}^2$ as a function of $x$
for five different values of $\chi$. In this figure the curves
$1-5$ from the bottom correspond to $\chi=0, 0.05, 0.10, 0.15$ and
$0.19$ respectively ($y \equiv v_{st}^2$).}
\end{figure*}

\begin{figure*}[ptbh]
\includegraphics[scale=.5]{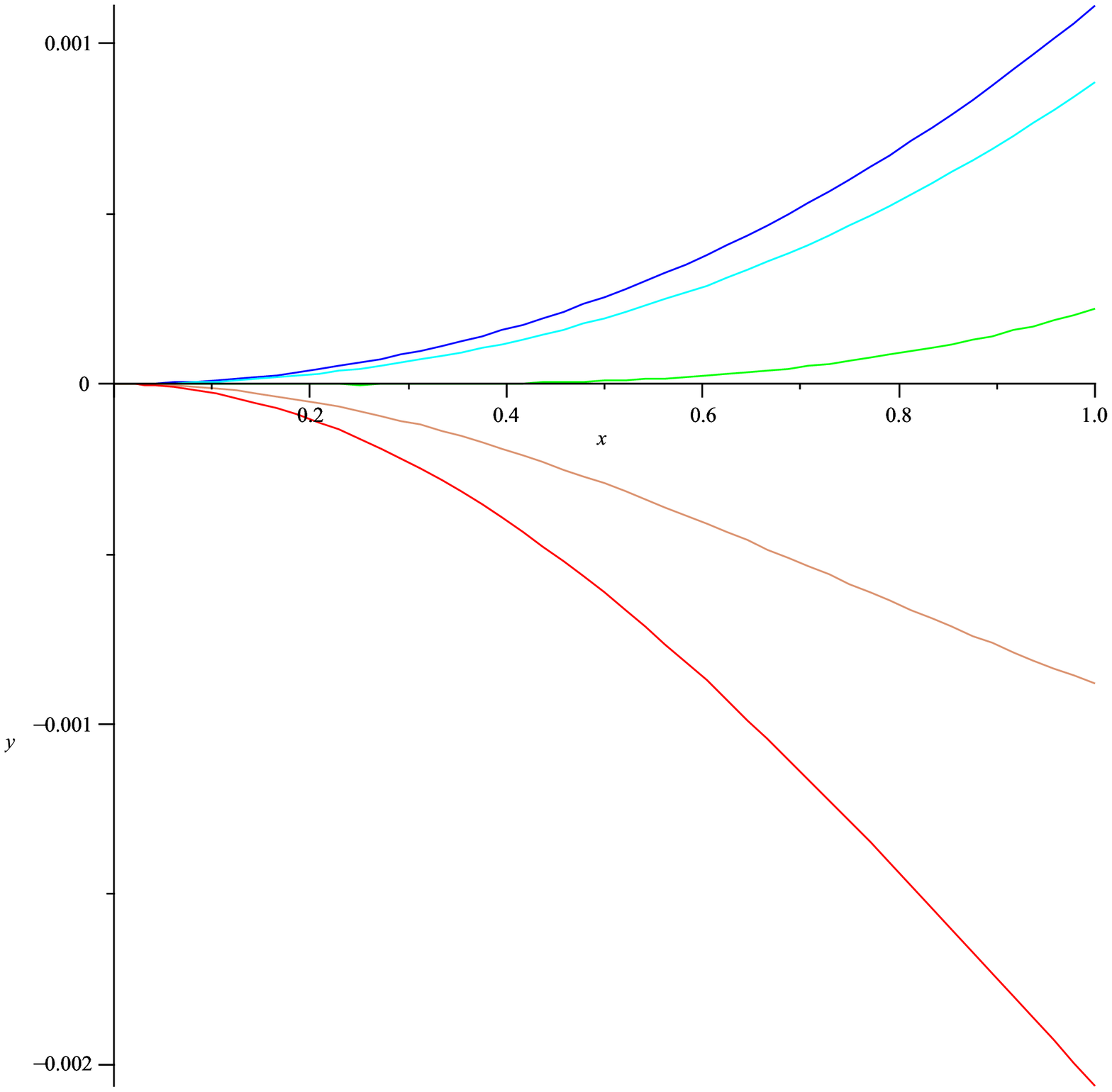}
\caption{Anisotropic parameter $\tilde{\Delta}$ for the first
model as a function of $x$ for five different values of $\chi$. In
this figure the curves $1-5$ from the top correspond to $\chi=0,
0.05, 0.10, 0.15$ and $0.19$ respectively ($y \equiv
\tilde{\Delta}$).}
\end{figure*}

\begin{figure*}[ptbh]
\includegraphics[scale=.5]{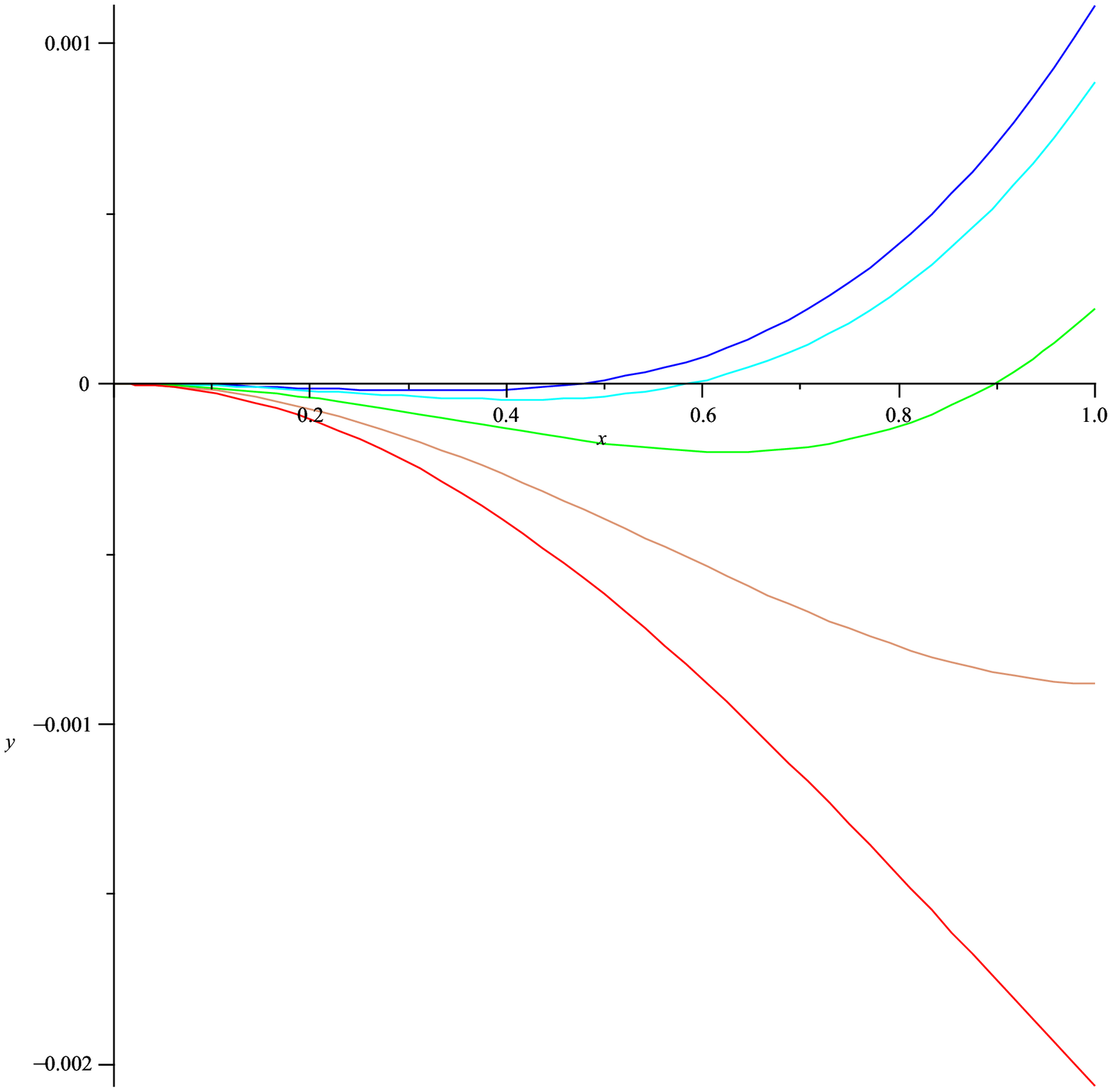}
\caption{Anisotropic parameter $\tilde{\Delta}$ for the second
model as a function of $x$ for five different values of $\chi$. In
this figure the curves $1-5$ from the top correspond to $\chi=0,
0.05, 0.10, 0.15$ and $0.19$ respectively ($y \equiv
\tilde{\Delta}$).}
\end{figure*}

\begin{figure*}[ptbh]
\includegraphics[scale=.5]{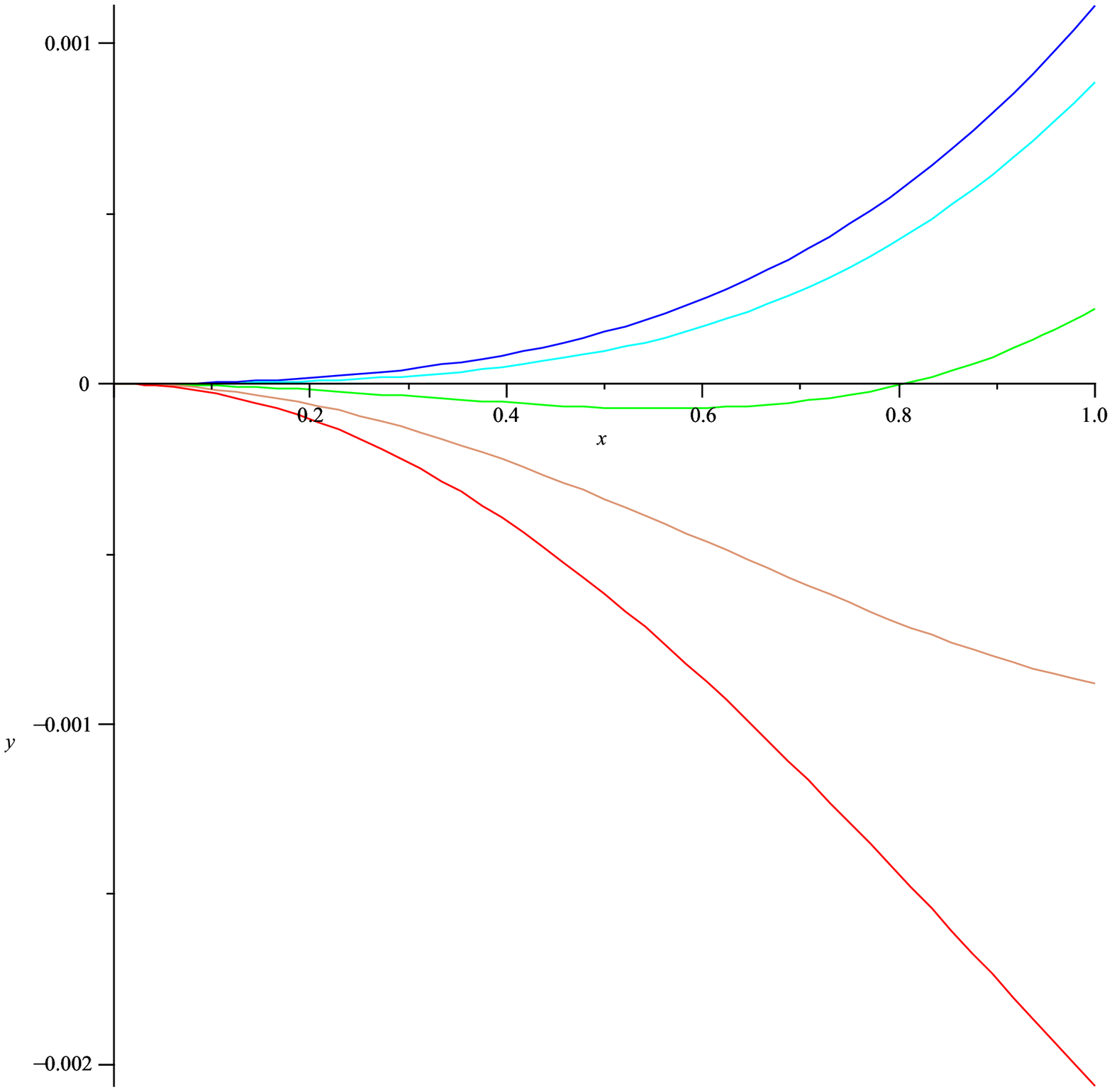}
\caption{Anisotropic parameter $\tilde{\Delta}$ for the third
model as a function of $x$ for five different values of $\chi$. In
this figure the curves $1-5$ from the top correspond to $\chi=0,
0.05, 0.10, 0.15$ and $0.19$ respectively ($y \equiv
\tilde{\Delta}$).}
\end{figure*}

\begin{figure*}[ptbh]
\includegraphics[scale=.5]{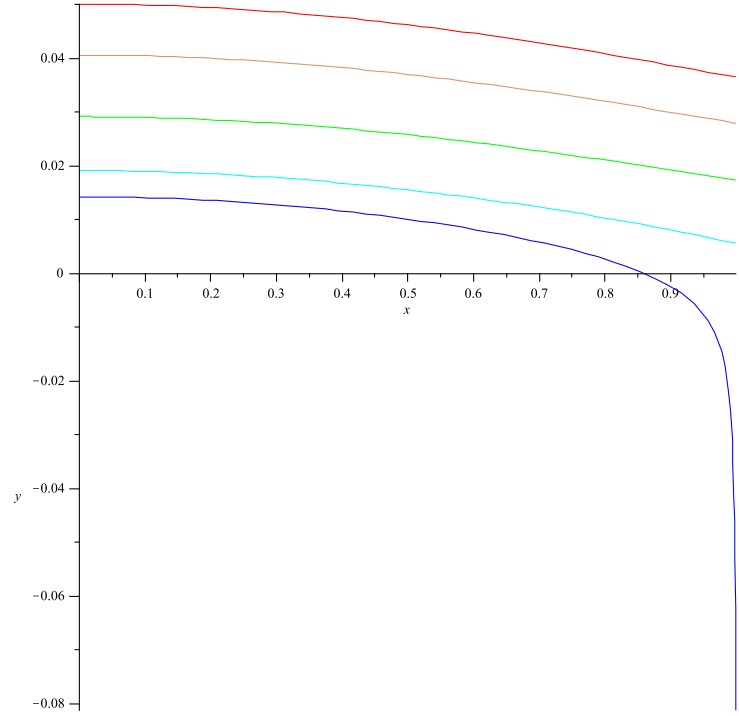}
\caption{Proper charge density $\tilde{\sigma}$ as a function of
$x\in[0,0.999]$ for five different values of $\chi$. In this
figure the curves $1-5$ from the bottom correspond to $\chi=0,
0.05, 0.10, 0.15$ and $0.19$ respectively ($y \equiv
\tilde{\sigma}$).}
\end{figure*}

\begin{figure*}[ptbh]
\includegraphics[scale=.5]{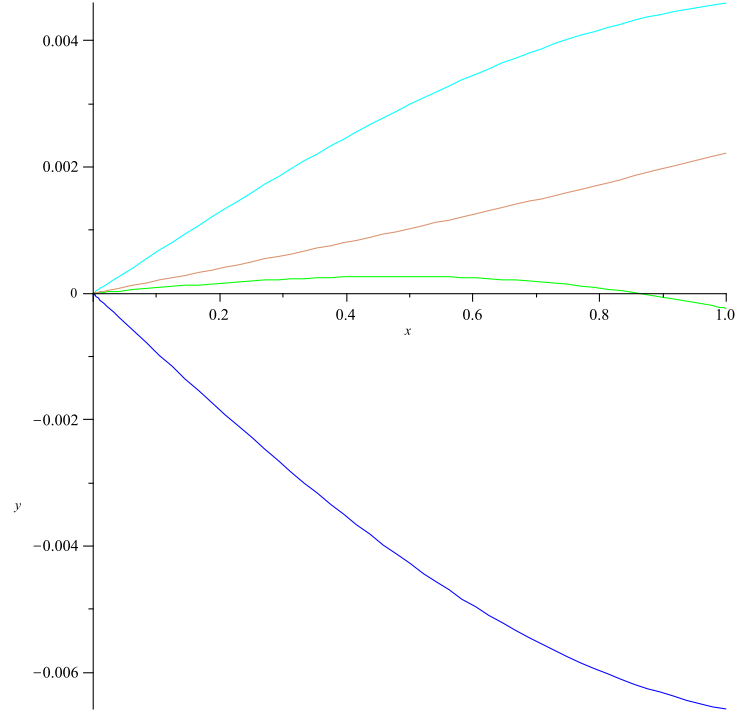}
\caption{Four different forces acting on fluid elements in static
equilibrium as functions of $x$ for $\mu$ = 0.147 and $\chi = 0$.
In this figure the curves 1-4 from the bottom (blue, green, tan,
cyan curves) correspond to $\tilde {F}_g$, $\tilde {F}_e$, $\tilde
{F}_a$, $\tilde {F}_h$, respectively. The dummy variable y on the
vertical axis represents any of these forces.}
\end{figure*}

\begin{figure*}[ptbh]
\includegraphics[scale=.5]{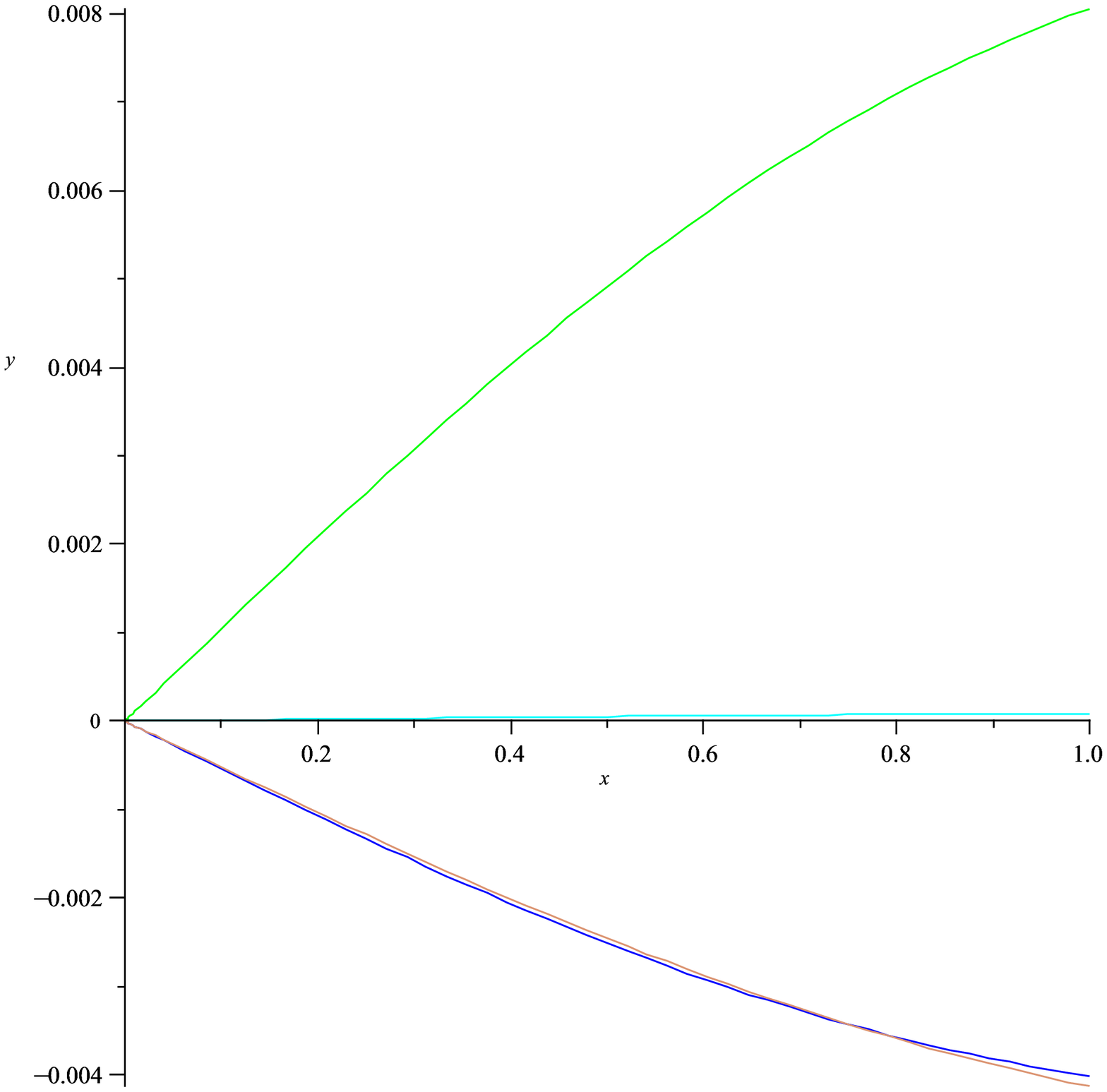}
\caption{Four different forces acting on fluid elements in static
equilibrium as functions of $x$ for $\mu$ = 0.147 and $\chi =
0.19$. In this figure the first (green) and second (cyan) curves
from the top correspond to $\tilde {F}_e$ and $\tilde {F}_h$,
respectively. The third (tan) and fourth (blue) curves intersect
at $x \approx 0.75$ and represent $\tilde {F}_a$ and $\tilde
{F}_g$, respectively. The dummy variable y on the vertical axis
represents any of these forces.}
\end{figure*}


\begin{thebibliography}{99}

\bibitem{iva1} B. V. Ivanov, Phys. Rev. D {\bf 65}, 104001 (2002).
\bibitem{sm1} R. Sharma and S. D. Maharaj, Mon. Not. R. Astron. Soc. {\bf 375}, 1265 (2007).
\bibitem{bon} B. W. Bonnor, Z. Phys. {\bf 160}, 59 (1960).
\bibitem{gs1} D. Horvat, S. Iliji{\'c} and A. Marunovi{\'c}, Class. Quantum Grav. {\bf 26}, 025003 (2009).
\bibitem{tm1} S. Thirukkanesh and S. D. Marahaj, Class. Quantum Grav. {\bf 25}, 235001 (2008).
\bibitem{kb} K. D. Krori and J. Barua, J. Phys. A: Math. Gen. {\bf 8}, 508 (1975).
\bibitem{lamu} K. Lake and P. Musgrave, Gen. Rel. Grav. {\bf 26}, 917 (1994).
\bibitem{bp1} O. Bertolami and J. P{\'a}ramos, Phys. Rev. D {\bf 72}, 123512 (2005).
\bibitem{jfr} M. Jamil, U. Farooq and M. A. Rashid, Eur. Phys. J. C {\bf 59}, 907 (2009).
\bibitem{pdl} J. Ponce de Le{\'o}n, Gen. Rel. Grav. {\bf 25}, 1123 (1993).
\bibitem{gmkps} We note that TOV equations have been used in the context of neutral
Chaplygin stars and wormwholes by V. Gorini, U. Moschella, A. Yu.
Kamenshchik, V. Pasquier and A. A. Starobinsky, Phys. Rev. D {\bf
78}, 064064 (2008); and V. Gorini, A. Yu. Kamenshchik, U.
Moschella, O. F. Piattella, and A. A. Starobinsky, Phys. Rev. D
{\bf 80}, 104038 (2009).
\bibitem{jnvc} G. J. G. Junevicus, J. Phys. A: Math. Gen. {\bf 9}, 2069 (1976).
\bibitem{dlpa} M. S. R. Delgaty and K. Lake, Comput. Phys. Commun. {\bf 115}, 395 (1998).
\bibitem{bhna} J. Burke and D. Hobill, arXiv: 0910.3230 (2009).
\bibitem{hmar} T. Harko and M. K. Mak, Annalen Phys. {\bf 11}, 3 (2002)
\bibitem{ahn} H.  Abreu, H. Hernandez and L. A. Nunez, Class. Quantum Grav. {\bf 24}, 4631 (2007).
\bibitem{bvhk} A. Balakin, J. W. van Holten and R. Kerner, Class. Quantum Grav. {\bf 17}, 5009 (2000).
\bibitem{vh} J. W. van Holten, arXiv: hep-th/0201083 (2002).
\bibitem{hh} J. B. Hartle and S. W. Hawking, Comm. Math. Phys. {\bf 26}, 87 (1972).
\bibitem{ml} P. Musgrave and K. Lake, Class. Quantum Grav. {\bf 12}, L39 (1995).
\bibitem{grw} R. Wald, {\it General Relativity} (Chicago: The University of Chicago Press, 1984).
\bibitem{mtw} C. W. Misner, K. S. Thorne and J. A. Wheeler, {\it Gravitation} (San Francisco: Freeman, 1973)
\bibitem{reml} S. Ray, A. L. Espindola, M. Malheiro and J. P. S. Lemos, Phys. Rev. D {\bf 68}, 084004 (2003).
\bibitem{ecsqs} R. P. Negreiros, F. Weber, M. Malheiro and V. Usov, Phys. Rev. D {\bf 80}, 083006 (2009).
\bibitem{madoha} M. K. Mak, P. N. Dobson and T. Harko, Europhys. Lett. {\bf 55}, 310 (2001).
\bibitem{champs} F. J. S\'{a}nchez-Salcedo, E. Mart\'{\i}nez-G\'{o}mez and J. Maga\~{n}a, arXiv: astro-ph.CO/1002.3145 (2010).
\bibitem{lobo} F. S. N. Lobo, Phys. Rev. D {\bf 75}, 024023 (2007).
\bibitem{hand} H. Andr{\'e}asson, Comm. Math. Phys. {\bf 288}, 715 (2009).

\end{thebibliography}
\end{document}